\catcode`\@=11                                   % To make protected \def's

%************************************************************
%*
%*            Font set-up
%*
%************************************************************

%************** 5-point fonts *******************************

\font\fiverm=cmr5                         % roman
\font\fivemi=cmmi5                        % math italic
\font\fivesy=cmsy5                        % math symbols
\font\fivebf=cmbx5                        % bold face

\skewchar\fivemi='177
\skewchar\fivesy='60

%************** 6-point fonts *******************************

\font\sixrm=cmr6                          % roman
\font\sixi=cmmi6                          % math italic
\font\sixsy=cmsy6                         % math symbols
\font\sixbf=cmbx6                         % bold face

\skewchar\sixi='177
\skewchar\sixsy='60

%************** 7-point fonts *******************************

\font\sevenrm=cmr7                        % roman
\font\seveni=cmmi7                        % math italic
\font\sevensy=cmsy7                       % math symbols
\font\sevenit=cmti7                       % italic
\font\sevenbf=cmbx7                       % bold face

\skewchar\seveni='177
\skewchar\sevensy='60

%************** 8-point fonts *******************************

\font\eightrm=cmr8                        % roman
\font\eighti=cmmi8                        % math italic
\font\eightsy=cmsy8                       % math symbols
\font\eightit=cmti8                       % italic
                       % slanted
\font\eightbf=cmbx8                       % bold face
                       % typewriter
                       % sans serif

\skewchar\eighti='177
\skewchar\eightsy='60

%************** 9-point fonts *******************************

\font\ninei=cmmi9
\font\ninesy=cmsy9

\skewchar\ninei='177
\skewchar\ninesy='60

%************** 10-point fonts ******************************

\font\tenrm=cmr10                         % roman
\font\teni=cmmi10                         % math italic
\font\tensy=cmsy10                        % math symbols
\font\tenex=cmex10                        % math extension
\font\tenit=cmti10                        % italic
\font\tensl=cmsl10                        % slanted
\font\tenbf=cmbx10                        % bold face
\font\tentt=cmtt10                        % typewriter
\font\tenss=cmss10                        % sans serif
\font\tensc=cmcsc10                       % small caps
\font\tenbi=cmmib10                       % bold math

\skewchar\teni='177
\skewchar\tenbi='177
\skewchar\tensy='60

\def\tenpoint{\ifmmode\err@badsizechange\else
    \textfont0=\tenrm \scriptfont0=\sevenrm \scriptscriptfont0=\fiverm
    \textfont1=\teni  \scriptfont1=\seveni  \scriptscriptfont1=\fivemi
    \textfont2=\tensy \scriptfont2=\sevensy \scriptscriptfont2=\fivesy
    \textfont3=\tenex \scriptfont3=\tenex   \scriptscriptfont3=\tenex
    \textfont4=\tenit \scriptfont4=\sevenit \scriptscriptfont4=\sevenit
    \textfont5=\tensl
    \textfont6=\tenbf \scriptfont6=\sevenbf \scriptscriptfont6=\fivebf
    \textfont7=\tentt
    \textfont8=\tenbi \scriptfont8=\seveni  \scriptscriptfont8=\fivemi
    \def\rm{\tenrm\fam=0 }%
    \def\it{\tenit\fam=4 }%
    \def\sl{\tensl\fam=5 }%
    \def\bf{\tenbf\fam=6 }%
    \def\tt{\tentt\fam=7 }%
    \def\ss{\tenss}%
    \def\sc{\tensc}%
    \def\bmit{\fam=8 }%
    \rm\setparameters\setbaselines\fi}

%************** 12-point fonts ******************************

\font\twelverm=cmr12                      % roman
\font\twelvei=cmmi12                      % math italic
\font\twelvesy=cmsy10       scaled\magstep1             % math symbols
\font\twelveex=cmex10       scaled\magstep1             % math extension
\font\twelveit=cmti12                            % italic
\font\twelvesl=cmsl12                            % slanted
\font\twelvebf=cmbx12                            % bold face
\font\twelvett=cmtt12                            % typewriter
\font\twelvess=cmss12                            % sans serif
\font\twelvesc=cmcsc10      scaled\magstep1             % small caps
\font\twelvebi=cmmib10      scaled\magstep1             % bold math

\skewchar\twelvei='177
\skewchar\twelvebi='177
\skewchar\twelvesy='60

\def\twelvepoint{\ifmmode\err@badsizechange\else
    \textfont0=\twelverm \scriptfont0=\eightrm \scriptscriptfont0=\sixrm
    \textfont1=\twelvei  \scriptfont1=\eighti  \scriptscriptfont1=\sixi
    \textfont2=\twelvesy \scriptfont2=\eightsy \scriptscriptfont2=\sixsy
    \textfont3=\twelveex \scriptfont3=\tenex   \scriptscriptfont3=\tenex
    \textfont4=\twelveit \scriptfont4=\eightit \scriptscriptfont4=\sevenit
    \textfont5=\twelvesl
    \textfont6=\twelvebf \scriptfont6=\eightbf \scriptscriptfont6=\sixbf
    \textfont7=\twelvett
    \textfont8=\twelvebi \scriptfont8=\eighti  \scriptscriptfont8=\sixi
    \def\rm{\twelverm\fam=0 }%
    \def\it{\twelveit\fam=4 }%
    \def\sl{\twelvesl\fam=5 }%
    \def\bf{\twelvebf\fam=6 }%
    \def\tt{\twelvett\fam=7 }%
    \def\ss{\twelvess}%
    \def\sc{\twelvesc}%
    \def\bmit{\fam=8 }%
    \rm\setparameters\setbaselines\fi}

%************** 14-point fonts ******************************

\font\fourteenrm=cmr12      scaled\magstep1             % roman
\font\fourteeni=cmmi12      scaled\magstep1             % math italic
\font\fourteensy=cmsy10     scaled\magstep2             % math symbols
\font\fourteenex=cmex10     scaled\magstep2             % math extension
\font\fourteenit=cmti12     scaled\magstep1             % italic
\font\fourteensl=cmsl12     scaled\magstep1             % slanted
\font\fourteenbf=cmbx12     scaled\magstep1             % bold face
\font\fourteentt=cmtt12     scaled\magstep1             % typewriter
\font\fourteenss=cmss12     scaled\magstep1             % sans serif
\font\fourteensc=cmcsc10 scaled\magstep2  % small caps
\font\fourteenbi=cmmib10 scaled\magstep2  % bold math

\skewchar\fourteeni='177
\skewchar\fourteenbi='177
\skewchar\fourteensy='60

\def\fourteenpoint{\ifmmode\err@badsizechange\else
    \textfont0=\fourteenrm \scriptfont0=\tenrm \scriptscriptfont0=\sevenrm
    \textfont1=\fourteeni  \scriptfont1=\teni  \scriptscriptfont1=\seveni
    \textfont2=\fourteensy \scriptfont2=\tensy \scriptscriptfont2=\sevensy
    \textfont3=\fourteenex \scriptfont3=\tenex \scriptscriptfont3=\tenex
    \textfont4=\fourteenit \scriptfont4=\tenit \scriptscriptfont4=\sevenit
    \textfont5=\fourteensl
    \textfont6=\fourteenbf \scriptfont6=\tenbf \scriptscriptfont6=\sevenbf
    \textfont7=\fourteentt
    \textfont8=\fourteenbi \scriptfont8=\tenbi \scriptscriptfont8=\seveni
    \def\rm{\fourteenrm\fam=0 }%
    \def\it{\fourteenit\fam=4 }%
    \def\sl{\fourteensl\fam=5 }%
    \def\bf{\fourteenbf\fam=6 }%
    \def\tt{\fourteentt\fam=7}%
    \def\ss{\fourteenss}%
    \def\sc{\fourteensc}%
    \def\bmit{\fam=8 }%
    \rm\setparameters\setbaselines\fi}

%************** Miscellaneous big fonts *********************

\font\seventeenrm=cmr10 scaled\magstep3          % roman
             % bold face

%************************************************************
%*
%*            Parameter initialization
%*
%************************************************************

\newdimen\rp@
\newcount\@basestretchnum
\newskip\@baseskip
\newskip\headskip
\newskip\footskip

% Routine to set page parameters

\def\setparameters{\rp@=.1em
    \headskip=24\rp@
    \footskip=\headskip
    \delimitershortfall=5\rp@
    \nulldelimiterspace=1.2\rp@
    \scriptspace=0.5\rp@
    \abovedisplayskip=10\rp@ plus3\rp@ minus5\rp@
    \belowdisplayskip=10\rp@ plus3\rp@ minus5\rp@
    \abovedisplayshortskip=5\rp@ plus2\rp@ minus4\rp@
    \belowdisplayshortskip=10\rp@ plus3\rp@ minus5\rp@
    \normallineskip=\rp@
    \lineskip=\normallineskip
    \normallineskiplimit=0pt
    \lineskiplimit=\normallineskiplimit
    \jot=3\rp@
    \setbox0=\hbox{\the\textfont3 B}\p@renwd=\wd0
    \skip\footins=12\rp@ plus3\rp@ minus3\rp@
    \skip\topins=0pt plus0pt minus0pt}

% Special routine to scale \baselineskip

\def\setbaselines{\maxdepth=4\rp@\baselinestretch=\@basestretchnum}

% The \baselinestretch command

\def\baselinestretch{\afterassignment\@basestretch\@basestretchnum}
\def\@basestretch{%
    \@baseskip=12\rp@ \divide\@baseskip by1000
    \normalbaselineskip=\@basestretchnum\@baseskip
    \baselineskip=\normalbaselineskip
    \bigskipamount=\the\baselineskip
        plus.25\baselineskip minus.25\baselineskip
    \medskipamount=.5\baselineskip
        plus.125\baselineskip minus.125\baselineskip
    \smallskipamount=.25\baselineskip
        plus.0625\baselineskip minus.0625\baselineskip
    \setbox\strutbox=\hbox{\vrule height.708\baselineskip
        depth.292\baselineskip width0pt }}

%************************************************************
%*
%*            Modifications to PLAIN.TEX
%*
%************************************************************

% Modifications to PLAIN routines to handle scaling of page parameters

\def\makeheadline{\vbox to0pt{\baselinestretch=1000
    \vskip-\headskip \vskip1.5pt
    \line{\vbox to\ht\strutbox{}\the\headline}\vss}\nointerlineskip}

\def\makefootline{\baselineskip=\footskip\line{\the\footline}}

\def\big#1{{\hbox{$\left#1\vbox to8.5\rp@ {}\right.\n@space$}}}
\def\Big#1{{\hbox{$\left#1\vbox to11.5\rp@ {}\right.\n@space$}}}
\def\bigg#1{{\hbox{$\left#1\vbox to14.5\rp@ {}\right.\n@space$}}}
\def\Bigg#1{{\hbox{$\left#1\vbox to17.5\rp@ {}\right.\n@space$}}}

% Modifications to PLAIN to handle bold math

\mathchardef\alpha="710B
\mathchardef\beta="710C
\mathchardef\gamma="710D
\mathchardef\delta="710E
\mathchardef\epsilon="710F
\mathchardef\zeta="7110
\mathchardef\eta="7111
\mathchardef\theta="7112
\mathchardef\iota="7113
\mathchardef\kappa="7114
\mathchardef\lambda="7115
\mathchardef\mu="7116
\mathchardef\nu="7117
\mathchardef\xi="7118
\mathchardef\pi="7119
\mathchardef\rho="711A
\mathchardef\sigma="711B
\mathchardef\tau="711C
\mathchardef\upsilon="711D
\mathchardef\phi="711E
\mathchardef\chi="711F
\mathchardef\psi="7120
\mathchardef\omega="7121
\mathchardef\varepsilon="7122
\mathchardef\vartheta="7123
\mathchardef\varpi="7124
\mathchardef\varrho="7125
\mathchardef\varsigma="7126
\mathchardef\varphi="7127
\mathchardef\imath="717B
\mathchardef\jmath="717C
\mathchardef\ell="7160
\mathchardef\wp="717D
\mathchardef\partial="7140
\mathchardef\flat="715B
\mathchardef\natural="715C
\mathchardef\sharp="715D

%************************************************************
%*
%*            Initialization
%*
%************************************************************

\def\err@badsizechange{%
    \immediate\write16{--> Size change not allowed in math mode, ignored}}

\baselinestretch=1000
\tenpoint

\catcode`\@=12                                   % Restore @ sign
% Routine to guarantee that this file is input only once
\catcode`\@=11
\expandafter\ifx\csname @iasmacros\endcsname\relax
    \global\let\@iasmacros=\par
\else  \immediate\write16{}
    \immediate\write16{Warning:}
    \immediate\write16{You have tried to input iasmacros more than once.}
    \immediate\write16{}
    \endinput
\fi
\catcode`\@=12

% Set up font size commands and \baselinestretch command
%\input iasfonts

% Some alternative font names
\def\rmb{\seventeenrm}

% Simple spacing commands
\def\singlespace{\baselineskip=\normalbaselineskip}
\def\halfspace{\baselineskip=1.5\normalbaselineskip}
\def\doublespace{\baselineskip=2\normalbaselineskip}

% Macros for references and abstracts

\def\AB{\bigskip\parindent=40pt
     \centerline{\bf ABSTRACT}\medskip\halfspace\narrower}
\def\AE{\bigskip\nonarrower\doublespace}
\def\nonarrower{\advance\leftskip by-\parindent
    \advance\rightskip by-\parindent}

% Useful commands

\def\boxit#1{\vbox{\hrule\hbox{\vrule\kern3pt
    \vbox{\kern3pt#1\kern3pt}\kern3pt\vrule}\hrule}}

% Special symbols
\def\hence{\leavevmode\hbox{\bf .\raise5.5pt\hbox{.}.} }

\def\dalemb#1#2{{\vbox{\hrule height.#2pt
    \hbox{\vrule width.#2pt height#1pt \kern#1pt \vrule width.#2pt}
    \hrule height.#2pt}}}
\def\gtorder{\mathrel{\raise.3ex\hbox{$>$}\mkern-14mu
          \lower0.6ex\hbox{$\sim$}}}
\def\ltorder{\mathrel{\raise.3ex\hbox{$<$}\mkern-14mu
          \lower0.6ex\hbox{$\sim$}}}

% For twoup output
\newdimen\fullhsize
\newbox\leftcolumn
\def\twoup{\hoffset=-.5in \voffset=-.25in
  \hsize=4.75in \fullhsize=10in \vsize=6.9in
  \def\fullline{\hbox to\fullhsize}
  \let\lr=L
  \output={\if L\lr
     \global\setbox\leftcolumn=\columnbox\global\let\lr=R \advancepageno
      \else \doubleformat \global\let\lr=L\fi
    \ifnum\outputpenalty>-20000 \else\dosupereject\fi}
  \def\doubleformat{\shipout\vbox{
    \fullline{\box\leftcolumn\hfil\columnbox}\advancepageno}}
  \def\columnbox{\leftline{\vbox{\makeheadline\pagebody\makefootline}}}
  \tolerance=1000 }
\twelvepoint
\doublespace
\overfullrule=0pt
{\nopagenumbers{
\rightline{~~~July, 2006}
\bigskip\bigskip
\centerline{\rmb Lower and Upper Bounds on CSL Parameters from}
\centerline{\rmb Latent Image Formation and IGM Heating}
\medskip
\centerline{\it  Stephen L. Adler}
\centerline{\bf Institute for Advanced Study}
\centerline{\bf Princeton, NJ 08540}
\medskip
\bigskip\bigskip
\leftline{\it Send correspondence to:}
\medskip
{\singlespace\leftline{Stephen L. Adler}
\leftline{Institute for Advanced Study}
\leftline{Einstein Drive, Princeton, NJ 08540}
\leftline{Phone 609-734-8051; FAX 609-924-8399;
email adler@ias.edu}}
\bigskip\bigskip
}}
\vfill\eject
\pageno=2
\AB
We study lower and upper bounds on the parameters for stochastic
state vector reduction, focusing on the mass-proportional continuous
spontaneous localization (CSL) model.
We show that the assumption that the state vector is reduced when a
latent image is formed, in photography or etched track detection, requires
a CSL reduction rate parameter $\lambda$ that is larger than conventionally
assumed by a
factor of roughly $2 \times 10^{9 \pm 2}$, for a correlation length $r_C$ of
$10^{-5}{\rm cm}$.
We reanalyze existing upper bounds on the
reduction rate and conclude that all are compatible with such
an increase in $\lambda$. The  best bounds that we have obtained come from
a consideration of heating
of the intergalactic medium (IGM), which shows that $\lambda$ can be at most
$\sim 10^{8\pm 1}$ times as large as the standard
CSL value, again for $r_C=10^{-5}{\rm cm}$.  (For both the lower and
upper bounds, quoted errors are not purely statistical errors, but rather
are estimates reflecting modeling uncertainties.)
We discuss  modifications in our analysis corresponding to a larger value
of $r_C$.
With a substantially enlarged rate parameter, CSL effects may be within range
of experimental detection (or refutation) with current technologies.
\AE
\bigskip\bigskip
\vfill\eject
\pageno=3
\centerline{\bf1.~~Introduction}
\bigskip
Stochastic modifications of the Schr\"odinger equation have been intensively
studied as models for objective state vector reduction [1].  As currently
formulated, the reduction rate parameters for these models are many decades
smaller than current experimental bounds, and will not be detectable in
planned nanomechanical and gravitational wave detector experiments [2].
In order to motivate further experimental searches for stochastic
modifications of Schr\"odinger dynamics, it is important to have
lower bounds on the stochastic model parameters, below which one can
assert that known measurement processes will not occur.  Setting
new, more stringent lower bounds, while at the same time reanalyzing
and improving upper  bounds, is the aim of this paper.

Conventional lower bounds on the stochastic parameters are based on an
idealized measurement model in which the experimenter reads out results
from the position of a macroscopic pointer.  In such measurements, the
detection and amplification processes needed to get a pointer readout
are necessarily linked.  Our focus in this paper is on a different type
of experiment, in which first a latent image is formed, either in a
photographic emulsion or a solid state track detector.  Only long after
latent image formation is amplification brought into play, in the
form of development of the photographic plate, or etching of the
track detector.  A qualitative discussion of latent photographic image
formation was given in 1993 by Gisin and Percival [3], who consider
detection to have occurred already at the microscopic level, {\it before}
the amplification associated with development.  To quote them, ``Of
particular importance is the study of the formation of the latent image in
photography, for this is not only the most common quantum detection
technology, but it also shows unequivocally that amplification up to the
macroscopic level is quite unnecessary for the formation of a permanent
classical record of a quantum event, by contrast with the example of
the pointer, which is so often used.''
In this paper we shall  put the discussion of Gisin and Percival on a
quantitative footing, within the framework of the
continuous spontaneous localization (CSL) model with mass-proportional
couplings.

This paper is organized as follows.  In Sec. 2 we give a quick review of the
CSL model with mass-proportional couplings, and in particular give
rate formulas
needed for the subsequent discussion.  In Sec. 3 we discuss the formation
of latent photographic images in the Mott-Gurney model,
and show that with standard parameter values, the
CSL model predicts a reduction rate that is a factor
of order  $\sim 2 \times 10^{9\pm 2}$
too slow, as compared with the estimated rate of latent image formation.
In Sec. 4, we make analogous (but cruder) estimates for solid state etched
track detectors, and again conclude that the usual CSL parameter
values cannot account for latent image formation.

In Sec. 5, we examine whether various
upper bounds on the CSL parameters allow a substantial  enlargement
in the reduction rate parameter.  Processes considered include
{}Fullerene diffraction, supercurrent persistence, proton decay, spontaneous
radiation from germanium,  cosmic IGM heating effects, and planetary heating.
Our conclusion in all cases not involving heating
(in some cases disagreeing with
previous analyses) is that an increase of $\lambda$ by a factor of  $10^{12}$
is allowed by experimental data.  Heating of the IGM places a
stronger constraint on $\lambda$, allowing an increase by a factor
of $\sim   10^{8\pm 1}$ over the standard value.
Although planetary heating nominally places a much more stringent bound, we
argue that competition with molecular collision effects, which are 28 orders
of magnitude larger and are strongly
dissipative, invalidates this bound.

In Sec. 6, we  discuss modifications in our analysis resulting from
changing the correlation
function $g(x)$, and from changing the value of the correlation
length $r_C$ from the standard value $10^{-5}$ {\rm cm} assumed
in CSL analyses.
In Sec. 7 we  discuss implications of our suggested new values for
the CSL parameters for experiments to directly test for CSL effects,
focusing on large molecule diffraction, superconductor current decay,
nanomechanical and gravitational wave detection experiments,
and the Collett-Pearle [4] proposal
to observe rotational Brownian diffusion.  We also briefly consider
the competition of molecular collisions with CSL effects.
{}Finally, in Sec. 8 we
discuss and summarize our results, and in particular, we note that
our key assumption, that latent image formation already constitutes
measurement, is subject to direct experimental test.
\bigskip
\centerline{\bf 2.~~The mass-proportional CSL model}
\bigskip
We begin by stating some standard formulas of the mass-proportional
CSL model, drawing heavily on the review of Bassi and Ghirardi [1].
The basic stochastic differential equation of the model is
$$d|\psi(t)\rangle=\left[ -{i\over \hbar}H dt + \int d^3x
(M(x)-\langle M(x)\rangle) dB(x)-{\gamma \over 2}
\int d^3x  (M(x)-\langle M(x)\rangle)^2 dt \right] |\psi(t)\rangle~,
\eqno(1)$$
with $dB(x)$ a Brownian motion obeying
$$dt dB(x)=0~,~~dB(x) dB(y)=\gamma\delta^3(x-y)dt ~~~,\eqno(2)$$
with $\langle M \rangle$ denoting the expectation of $M$ in
the state $|\psi(t)\rangle$, and with the operator $M(x)$ given by
$$M(x)= m_N^{-1}\int d^3 y g(x-y) \sum_s m_s N_s(y)~~~.\eqno(3)$$
In Eq.~(3) the sum extends over particle species $s$ of mass $m_s$ and
with number density operator $N_s(y)$, while
$m_N$ is the mass of the nucleon and $g(x)$ is a spatial correlation function
conventionally chosen as
$$g(x)=\left({\alpha \over 2 \pi}\right)^{3/2}e^{-(\alpha/2)x^2}~,~~
\int d^3x g(x) =1~~~.\eqno(4)$$
(In Sec. 6 we will show that changing the functional form
of the correlation function does not substantially alter our conclusions.)
This correlation function can be interpreted as the functional
``square root'' of
the correlation function for a noise variable that couples locally
to the mass density $\sum_s m_s N_s(y)$.  That is, writing
$dC(y)=\int d^3 x g(x-y) dB(x)$, which has the spatial correlation
function $dC(y)dC(z)=\gamma dt \int d^3x g(x-y)g(x-z)$, the noise term
in Eq.~(1) takes the locally coupled form
$\int d^3 y dC(y) m_N^{-1} \sum_s m_s (N_s(y)-\langle  N_s(y)\rangle)$.
Thus, the basic parameters of the model are the strength of the Brownian
process $\gamma$, and the correlation function width parameter $\alpha$.

It is convenient to introduce two further parameters that are defined in
terms of $\gamma$ and $\alpha$.  Since $\alpha$ has the dimensions of
inverse squared length, we define a correlation length $r_C$ (sometimes
denoted by $a$ in the CSL literature) by
$$r_C=(\alpha)^{-1/2}~~~,\eqno(5a)$$
which is conventionally assumed to take the value of $r_C=10^{-5}$ cm.
Additionally, we introduce a rate parameter $\lambda$ defined by
$$\lambda=\gamma \left( {\alpha\over 4 \pi}\right)^{3/2}
=\gamma/(8 \pi^{3/2} r_C^3)~~~,\eqno(5b)$$
which is conventionally assumed to take the value
$\lambda=2.2 \times 10^{-17} \,
{\rm s}^{-1}$, giving $\gamma$ the value $10^{-30} {\rm cm}^3
{\rm s}^{-1}$.

We can now state two key rate formulas that we will need for the subsequent
analysis.  According to Eq.~(8.15) of Bassi and Ghirardi [1], the
off-diagonal coordinate space density matrix
element $\langle \ell|\rho|0 \rangle$ for a single nucleon
approaches zero exponentially with a reduction rate $\Gamma_R$ given
by
$$\Gamma_R=\lambda\big(1-e^{-\ell^2/(4 r_C^2)}\big)~~~,\eqno(6a)$$
which for $\ell$ comparable to or larger than $r_C$, can be approximated as
$$\Gamma_R \simeq  \lambda~~~.\eqno(6b)$$
{}For $n$ nucleons within a radius smaller than the correlation length,
this rate is multiplied by $n^2$; for $N$ groups of nucleons separated by
more than the correlation length, this rate is multiplied by $N$, and for
particles of mass $m_p$ the rate is multiplied by $(m_p/m_N)^2$, giving
$$\Gamma_R \simeq \lambda n^2 N (m_p/m_N)^2~~~.\eqno(6c)$$
The formulas of Eqs.~(6a-c) will be the basis of our reduction time estimates
for latent image formation.  Before proceeding further, we note
that there are other natural definitions of a reduction rate; for
example, if it is defined by the rate of approach to zero of the variance
in the coordinate, as in Adler [1], rather than the rate of vanishing of the
off-diagonal density matrix element, then $\Gamma_R$ is double that given
by Eqs.~(6a-c).

A second important formula gives the rate of secular center-of-mass
energy gain, as
a result of the Brownian process, for a body comprised of
a group of particles of total mass
$M$.    This is given by
the formula (Bassi and Ghirardi [1], Pearle and Squires [5], Adler [6])
$${dE \over dt}= {3 \over 4}\lambda  {\hbar^2 \over r_C^2} {M \over m_N^2}
~~~.\eqno(7)$$
This formula will be used to set upper bounds on the reduction rate parameter
$\lambda$.
\bigskip

\centerline{\bf 3.~~Latent image formation in photography}
\bigskip
The case of latent image formation in photography is of particular
interest because the photographic process has been intensively studied,
both theoretically and experimentally.  Survey accounts of what is known
are given in the books of Mott and Gurney [7] and
Avan et al [8], and in the review articles of Berg [9] and of
Hamilton and Urbach [10].

A photographic
emulsion consists of grains of silver halide suspended in gelatine. The
typical grain size is between $1/10$ of a micron
and a few microns, or in other words,
from $10^{-5}$ cm to a few times $10^{-4}$ cm in diameter.  Grains are
typically spaced around 1 to 2.5 microns apart in the gelatine. The basic
microscopic theory originated by Gurney and Mott [11] is illustrated
graphically on p 75 of ref [8], for a typical AgBr emulsion.  The steps
as envisaged by Gurney and Mott are as follows.  First a photon is absorbed
in the grain, giving rise to an electron and a hole.  The electron gets
trapped on the surface of the grain, and the hole produces a neutral Br
which is absorbed on the surface.  An interstitial
ion of ${\rm Ag}^+$ then diffuses to the trapped electron and forms a neutral
silver atom. A second photon is absorbed, giving rise to another electron
and hole; the hole produces a second neutral Br which joins with the first
to produce a ${\rm Br}_2$ molecule, which eventually diffuses
out of the grain
into the gelatine, while the electron combines with the
neutral Ag on the surface to give an ion ${\rm Ag}^-$. (As shown in Fig. 5.4
of Hamilton and Urbach [10], the bromine that
has diffused into the gelatine typically moves up to around a micron away
from the grain; this will figure in our discussion of Sec. 6.)
{}Finally, a second
interstitial ${\rm Ag}^+$ diffuses to the surface to
join the ${\rm Ag}^-$, forming
a molecule ${\rm Ag}_2$.  This process is then repeated until a cluster of
free silver atoms (a silver speck)
has been formed on the surface of the grain, making
it developable.   As discussed in detail in ref [8], an alternative
microscopic model has been proposed by Mitchell,
in which crystalline imperfections play an important
role, and in which the order of steps differs from that in the theory of
Gurney and Mott.  However, such details are not relevant to the estimates
that we shall make, which only use the number of ions that move and the
distance that they travel, but not the order in which the motions take
place, details of trapping, etc.

There seems to be general agreement that around 3 to 6 silver
atoms are needed
to give a grain a 50\% probability of being developable, and about 30
silver atoms are needed for a grain to be certain to be developable.
In using grains to define a particle track in a 100 micron width emulsion,
a few grains developable by random
processes are encountered, so we shall assume that a minimum number of
about $N=20$ developable grains is required to define a track.  Assuming
that 30 Ag and Br atoms (with respective atomic weights of 108 and 80)
move a distance greater than or of order $r_C$ in the ionic
step of latent image formation, we have for the number of nucleons that
move $n=30(108+80)=5640$.  So from Eq.~(6c), we have for the reduction
rate $\Gamma_R$ for the process of latent image
formation producing the track,
$$\Gamma_R=\lambda n^2 N= 2.2 \times 10^{-17}{\rm s}^{-1}\times
5640^2\times 20
=1.3 \times 10^{-8} {\rm s}^{-1} ~~~.\eqno(8)$$

On the other hand, Mott and Gurney [7] estimate in their Table 28 the
rate of growth of a silver speck.  Scaling their room temperature
example to a single speck of radius $0.2 \times 10^{-5}
{\rm cm}$, which fits comfortably
within a grain of diameter $10^{-5}{\rm cm}$,
one finds a rate for accumulating 30 silver atoms of
around $30 {\rm s}^{-1}$.
This rate should be considered uncertain by two orders of magnitude
in either direction; for example, using the ion step time of
$1/25,000$s discussed on p 248 of Mott and Gurney,
the rate to accumulate 30 silver atoms would be
about $1000 {\rm s}^{-1}$, while applying the results of Table
28 to a speck of radius $0.2 \times 10^{-7}{\rm cm}$, corresponding to
just a few silver atoms, would give a rate to accumulate 30 atoms
of $0.3 {\rm s}^{-1}$. Note that since the accumulation of silver atoms on
a speck is a sequential process, the time for building up the speck
can be much longer than the duration of the exposure that
initially produces electrons and holes in the interior of the grain.

If we identify the time to form a developable speck with the
total time to form the latent image, we see that a reduction rate
of $3 \times 10^{1 \pm 2} {\rm s}^{-1}$ is required for reduction
to be completed
during latent image formation. Thus, based on this
estimate, the rate of Eq.~(8) is too small by a factor of $2 \times
10^{9\pm 2}$; in other words, the CSL rate parameter
$\lambda$ would have to be increased by this factor to account for latent
image formation (for  $r_C$ fixed at $10^{-5}$cm).

Let us now address a possible objection to this conclusion,
that we have considered only the mass motion associated with registration
of the latent image, but have neglected possible motions of environmental
particles induced by the registration process.  The first thing to be said
is that photographic emulsions  work
perfectly well in vacuo, so it suffices to consider just the motions of
particles within the emulsion.  When a photon is absorbed
by an atom, the atom will recoil, potentially altering the phonon
distribution, and thus displacing all the atoms in the detector.  For a
photon of energy a few eV  absorbed by a silver halide molecule,
this recoil will have an energy $E_R$ of order
$10^{-10}{\rm eV}$,  which is much
less than the typical Debye temperature energy $E_D \sim 10^{-2}{\rm eV}$.
Thus the parameter $E_R/E_D$ governing the
probability of a phonon-free absorption is of order $10^{-8}$, and so the
recoil will be absorbed, with  probability $1-{\rm O}(10^{-8})$,
by a translation of the whole detector, without the emission of phonons.
Hence for relevant elapsed times $t$, the
recoil distance
$\ell$ will be much smaller than the correlation length, and we can use
the expansion of Eq.~(6a) given in Eq.~(9) below.   Given a total
detector mass of $Nnm_N$, the recoil distance will be
$\ell=pt/(Nnm_N)$, with $p$ the incident photon momentum, and so
the reduction
rate associated with detector recoil will be $\Gamma_R =[\lambda/(4N)]
[pt/(m_N r_C)]^2$.  Notice that the factor $n^2$ has canceled out, and
the rate is proportional to the inverse of the emulsion size, as reflected
in the factor $N^{-1}$.  For a $3 {\rm eV}$ incident photon, a correlation
length $r_C=10^{-5} {\rm cm}$, and an elapsed time
$t$ of $1/30 {\rm s}$, this estimate gives $\Gamma_R \sim 0.5 N^{-1} \times
10^{-6} {\rm s}^{-1}$ for the standard CSL value of $\lambda$.
For a minimal detector with
$N=20$ we obtain a reduction rate
about a factor of two larger than that estimated in Eq.~(8), but for a
one cubic centimeter emulsion, with $\sim 10^{24}$ nucleons, the effective
$N$ in the recoil estimate is $10^{15}$, and the recoil induced reduction
rate is negligible.

Once the incident photon has been absorbed, the remaining steps
in latent image formation involve transport and
diffusion of electrons, holes, or
silver and bromine ions.  The electron liberated from a
silver halide molecule
by the absorbed photon will initially have a velocity larger than the
mean thermal velocity, and will
emit phonons until it slows down to thermal velocity.
This effect is calculated in Appendix D, where we estimate that it leads to
a reduction rate $\Gamma_R \sim  10^{-17} {\rm s}^{-1}$,
nine orders of magnitude smaller than the result of Eq.~(8).
Assuming that interstitial ions are already thermalized, and that they have a
dispersion relation of the form $E=p^2/2 m^*$ with $m^*$ an effective
mass, the
corresponding phonon emission reduction rates due to ionic motions are
much smaller, because
only the tail of the ionic thermal momentum distribution is kinematically
allowed to emit phonons.  Once thermal equilibrium is established, only
the difference between the thermal motions in superposed states in the wave
function can contribute to reduction. This difference should be smaller than
twice the thermal motion in one of the states of the superposition,
and an estimate given in Appendix D
shows that this bound on the thermal contribution to the reduction rate
is similar in magnitude to the result of Eq.~(8).
{}Finally, the motion of a silver
(or bromine) ion from the interior of a grain to the surface requires a
compensating back motion of the whole emulsion if no momentum is transferred
to the surroundings.
For example, if a silver ion diffuses a distance
$r_C$, to keep the center of mass of the emulsion fixed, it must translate
by $108 r_C/(Nn)$, leading to a reduction rate $\Gamma_R \sim N^{-1} \times
10^{-13} {\rm s}^{-1}$ for the standard CSL value of $\lambda$.
This is negligibly small even for $N=1$, that is,
for an emulsion consisting of a single grain.
In sum, these estimates suggest that collective motions of the
atoms in the emulsion, induced by the process of latent image formation,
do not alter our conclusion that the standard CSL value of $\lambda$ is not
large enough to account for latent image formation.

\bigskip
\centerline{\bf 4.~~Latent image formation in etched track detectors}
\bigskip
In this section we turn to the consideration of a second type of latent
image formation, in which etchable defects are formed in solid state
detectors.  Using related parameters,
we will also estimate the reduction time associated with
electronic motions resulting from passage of the charged particle through
the detector.

Detailed accounts of etched track detectors are given in the books of
Durrani and Bull [12] and Avan et al [8], from which we draw the
needed parameters.  Passage of a charged particle results in
a cylindrical region of lattice distortion around the charged particle
path, with a typical radius of 6 to 12 nm (i.e., 60 to 120 \AA,
or 0.6 to 1.2 $\times 10^{-6}$ cm), which we take as giving the displacement
$\ell$ in Eq.~(6a).  Since this $\ell$ is considerably smaller
than $r_C$, we  expand the exponential to get the associated reduction rate
$$\Gamma_R=\lambda n^2 N \ell^2/(4 r_C^2)~~~,\eqno(9)$$
where as before, $n$ is the number of nucleons within a correlation length,
and $N$ is the number of correlation-length groups of nucleons in the
particle track.  Let us estimate the length of a short ionic track length
to be 10 microns, giving $N=100$.  For $n$, we take the number of nucleons
in a cylinder of diameter $\ell= 10^{-6}$ cm, and length
$10^{-5}$ cm, assuming a density of $10^{24}$ nucleons per cubic centimeter.
This estimate gives $n\simeq 3 \times 10^7$.  Thus,
with the standard value $\lambda=2.2 \times 10^{-17}{\rm s}^{-1}$, we
get a reduction
rate of  $\Gamma_R \simeq 7\times 10^{-3} {\rm s}^{-1}$.  Various numbers for the
time $t$ for a track to form are given in the two books cited.  A minimum
for $t$ is given by the lattice vibration time of $10^{-13}$s; according
to  Table VIII.3 of Avan et al, establishment of thermal equilibrium
requires a time in the range of $10^{-12}$s to $10^{-9}$s, and establishing
chemical equilibrium requires a time of a few times $10^{-8}$ s to
much longer, depending on the material.
Thus, these time estimates indicate that the
standard value of $\lambda$ gives a reduction rate that is too small by
a factor of $10^{11}$ to $10^{14}$, if thermal equilibrium is taken as
the criterion for latent image formation, and a factor of
around $5 \times 10^{9}$ or less, if chemical equilibrium
is used as the criterion.

Let us next address a potential objection to this conclusion, and to the
similar one reached in Sec. 3
in the case of photographic emulsion grains,  that we
have neglected the mass transport associated with the ionization produced
by passage of the charged particle.  However, the initial ionization
consists almost entirely of the motion of electrons, and according to Katz
and Kobetich [13], the bulk of the energy deposited within $20$\AA~ of the
track comes from $\delta$-rays  with energy less than $0.1$keV.
As one might expect, these electrons have ranges of order $40$\AA~ or less,
so the geometry of electron displacements is essentially the same as
the geometry of nucleonic displacements.
Since the electron-associated reduction rate is smaller than that
calculated above by a factor $(m_e/m_N)^2$, the electrons that are initially
ionized in the course of track formation give a negligible
correction to the estimate made above.

\bigskip
\centerline{\bf 5.~~Upper bounds on the CSL parameter}
\bigskip
In the previous two sections, we have concluded that if state vector
reduction occurs when a latent image is formed, the reduction rate parameter
$\lambda$ must be much larger than conventionally assumed.  We turn now
to an examination of whether such an enlarged $\lambda$ is allowed by
various experimental constraints.  We continue to assume for this
discussion that the correlation length $r_C$ is $10^{-5}$cm, an assumption
to which we shall return in Sec. 6.
\bigskip
\centerline{\bf 5A.~~Fullerene diffraction experiments}
\bigskip
We begin our survey of experimental constraints with the fullerene
diffraction experiments of Arndt et al [14], Nairz, Arndt and Zeilinger [15],
and Nairz et al [16], which we previously discussed in Sec. 6.5 of Adler [1].
In these experiments, molecules
with $n\sim 10^3$ nucleons are observed to produce
interference fringes when diffracted through gratings with spacings of
 1 to $2.5 \times 10^{-5}$cm, of order the correlation length $r_C$.
{}From Eq.~(6c), we get a reduction rate with the standard
$\lambda$ of $\Gamma_R=2.2 \times 10^{-17}{\rm s}^{-1}
\times  1000^2 \simeq 2 \times 10^{-11} {\rm s}^{-1}$. Since the beam transit time in these
experiments is of order $10^{-2}$s, the interference fringes will not be
spoiled provided that $\Gamma_R < 10^2 {\rm s}^{-1}$, permitting the
rate parameter $\lambda$ to be up to $5 \times 10^{12}$ times as large as the
standard value.
\bigskip
\centerline{\bf 5B.~~Decay of supercurrents}
\bigskip
The decay of supercurrents in the mass-proportional CSL model has
been calculated by Buffa, Nicrosini and Rimini [17], as also reported
in Rimini [18], extending to the CSL case earlier work by Rae [19].
Equation (5.9) of ref [17] gives a fractional supercurrent decay rate, using
the standard
CSL parameters, of $5.1\times 10^{-27} {\rm s}^{-1}$.  Since the current
experimental limit is [20] about $10^{-5} {\rm year}^{-1}=3 \times
10^{-13}{\rm s}^{-1}$, the rate parameter $\lambda$ is permitted to
be up to $10^{14}$ times as large as the standard value. (This limit will
be increased when recombination processes leading to the formation
of Cooper pairs are taken into account, since these are not included in
the calculations of refs. [17], [18], and [19].)

The review of Bassi and Ghirardi [1] quotes in detail only the numbers
obtained by Rae, although referring to the later work of refs [17] and [18].
Rae assumes a fractional decay rate given by  $\lambda$,
for which he takes the value $10^{-15}{\rm s}^{-1}$, as opposed to the CSL
value of $2.2 \times 10^{-17}{\rm s}^{-1}$.
The CSL model calculation in refs [17] and [18] decreases the estimate
given by $\lambda$ by
a factor of $(r_Ck_F)^{-1}\simeq 0.6 \times 10^{-3}$ (with $k_F$ the
{}Fermi momentum) arising from
indistinguishability of electrons, and a factor $(m_e/M_N)^2
\simeq 0.3 \times 10^{-6}$, arising from the reduced stochastic
coupling of the electron in the mass-proportional scheme.  Assuming that
each spontaneous localization breaks a Cooper pair, and that Cooper
pairs are not recreated, this gives
an overall supercurrent decay rate of
$2.2 \times 10^{-17}{\rm s}^{-1} \times 0.2 \times 10^{-9} = 4.4 \times
10^{-27}{\rm s}^{-1}$, essentially the same number given in ref [17].

The indistinguishability factor $(r_Ck_F)^{-1}$ can be understood
heuristically from the fact that localization within a radius $r_C$ involves
a momentum transfer $ \delta k \sim 1/r_C$, and so is forbidden by the
exclusion principle except for electrons within $\delta k$ of the Fermi
surface, which are a fraction of order $\delta k/k_F$ of all electrons.
Since for temperatures well below the critical temperature all electrons
in the Fermi sea contribute to the supercurrent, and not just those
near the Fermi surface, spontaneous localization
can thus affect only a fraction
$(r_Ck_F)^{-1}$ of the electrons in the supercurrent, and hence this factor
appears in the supercurrent decay rate.
\bigskip
\centerline{\bf 5C.~~Excitation of Bound Atomic and Nuclear Systems}
\bigskip
Pearle and Squires [5] have estimated the rate of bound state excitation
in the mass-proportional CSL model and applied it to various
physical systems.  Their Eq.~(12) gives  for the rate $\dot P$ of internal
excitation of atoms
$$\dot P\sim \lambda (m_e/m_N)^2 (a_0/r_C)^4~~~,\eqno(10)$$
with $a_0$ the atomic radius $\sim 10^{-8}{\rm cm}$,
which with the standard value of $\lambda$ gives
$\dot P \sim 0.7 \times 10^{-35} {\rm s}^{-1}$.  Requiring that cosmic
hydrogen not dissociate during the lifetime of the universe of $4 \times
10^{17}$s gives an upper bound on $\lambda$ of $4 \times 10^{17}$ times
the standard value.

Also in ref [5], Pearle and Squires apply an analogous formula to
proton decay, with $a_0$ now a nuclear radius of $10^{-13}$ cm, and
with $m_e$ replaced by the quark mass $m_q$.  Using the constituent
quark mass $m_q \sim m_N/3$, this gives
$\dot P \sim 10^{-50} {\rm s}^{-1}$, while using current quark masses
of around 10 MeV gives $\dot P \sim 10^{-53} {\rm s}^{-1}$.  Requiring
$\dot P$ to be smaller than the  proton decay rate, which is known to be
less than $10^{-33} {\rm year}^{-1}=0.3 \times 10^{-40} {\rm s}^{-1}$,
allows $\lambda$ to be up to $\sim  10^9$ times the standard value if
constituent quark masses are used in the estimate, and up
to $\sim 10^{13}$ times the
standard value if current quark masses are used.  However,
these limits are overly restrictive because the formula of Pearle and Squires
assumes that no selection rules are at work.  This is true in the case of
atomic dissociation, but not in the case of decay of a proton.  Because
free quarks are confined, and because fermion number is conserved modulo 2,
allowed proton decay modes must contain at least one lepton, which is not
originally present as a constituent of the proton.  Hence the proton
decay rate,  which involves a wave function overlap squared,
must be further suppressed by $(m_N/\Lambda)^4$, with $\Lambda$ the
minimum scale at which physics beyond the standard model is found.  Making
the very conservative estimate $\Lambda \sim 250 {\rm GeV}$ (the electroweak
scale) gives an additional suppression of $3 \times 10^{-10}$; hence
proton decay allows $\lambda$ to be up to at least $10^{18}$ times
the standard value, using the constituent quark mass estimate.

Yet another application of the excitation rate formula is given by
Collett and Pearle [4], to an experiment in which the rate of spontaneous
11  keV photon emission from germanium, as monitored in 1 keV bins,
has been bounded by
0.05 pulses/(keV kg day).  The assumption here is that nuclear excitation
will have a high probability of knocking out a 1s electron, with resulting
emission of an 11keV photon as the remaining electrons cascade downward
to fill the vacant orbit.  To get an estimate we use
Eq.~(10) with $m_e$ replaced by
$m_N$, and  $a_0$ taken as the nuclear radius given by
$1.4 \times 10^{-13} A^{1/3}$
cm, with $A$ the atomic number. Using  $A\simeq 73$ for germanium, which has
$8.3 \times 10^{24}$ atoms/kg, and equating $\dot P$ to the rate limit
for the first 1keV bin, gives the bound $\lambda <6 \times  10^{-3}
{\rm s}^{-1}$,
which is $\sim 3 \times 10^{14}$ times as large
as the standard CSL value of $\lambda$.  (This bound is a factor of $10^3$
higher than the one quoted by Collett and Pearle [4]; where they get
$\lambda^{-1}r_C^4> 2 \times 10^{-15}  {\rm cm}^4 {\rm s} $, our evaluation
from their numbers gives $\lambda^{-1}r_C^4 >2 \times 10^{-18}
{\rm cm}^4  {\rm s} $.)  Yet another analysis of nuclear excitation
has recently been given by Pearle [4], who used the
CSL model to estimate the rate of quadrupole
excitation of a germanium nucleus to its first excited state. Pearle shows
that for this process,
the experimental data on photon emission require $\lambda$ to be
less than $\sim 10^{14}$ times its standard value.
\bigskip
\centerline{\bf 5D.~~Radiation by Free Electrons}
\bigskip
{}Fu [21] has calculated the rate of radiation of photons of energy $k$ by
free electrons that is
induced by the stochastic term in the Schr\"odinger equation.  Including
the factor $(m_e/m_N)^2$ called for in the mass-proportional coupling case,
his result becomes
$${d\Gamma(k)\over dk}={e^2 \lambda
\over 4 \pi^2 r_C^2 m_N^2 k}~~~.\eqno(11)$$
An alternative derivation of this result is given in Appendix A,
obtained by calculating the mean squared acceleration
produced by the stochastic term, and then using this in the
standard formula for the power radiated by an accelerated charge.
In evaluating his result numerically, Fu takes $e^2=1/137.04$, whereas the
standard Feynman rules that he uses require the identification
$e^2/(4 \pi)=1/137.04$. He also uses the
value $\lambda=10^{-16}{\rm s}^{-1}$,
so to correct for his evaluation of $e^2$ and to correspond to the CSL
value of $\lambda$, we multiply the rate result of his Eq.~(4.1) by
$4\pi/4.5=2.8$. Comparing with the experimental
bound on photon emission from germanium
quoted above, this gives $\lambda<1.7\times 10^{-11}{\rm s}^{-1}$, which
is only $\sim  10^6$ times as large as the standard CSL value of
$\lambda$.

However, in the special case of mass-proportional coupling that we are
considering here, this bound is significantly modified.
Although the valence electrons in germanium are quasi-free, they are still
charge neutralized by the core consisting of the inner electrons and nucleus.
The core has a center of mass motion with a characteristic lattice
vibration period of order $2 \times 10^{-13}{\rm s}$, much longer than the
$4 \times 10^{-19}{\rm s}$  period of an emitted 11 keV photon, so the
core motion should give at most an order $10^{-6}$ correction to
the rate of 11 kilovolt
photon emission by the core induced by the stochastic term
in the Schr\"odinger equation.   In the mass-proportional
coupling CSL model, since the core radius is much less than the
correlation length,
the core can be treated as a point particle with
the stochastic term acting only
on its center of mass.  Moreover,
the acceleration of the core induced by the stochastic term will be the
same as that of electrons (the $m^{-1}$ factor relating acceleration to
force is canceled by the $m$ factor in the coupling to the noise), and
so the core motion will tend to cancel the radiation field produced by the
electron motions.  The cancellation in the far-zone radiation field is
incomplete  because of two types of corrections.  The first are
order $a_0/r_C$ corrections, reflecting the variation of
the correlation function over the germanium atom radius
$a_0 \sim 10^{-8} {\rm cm}$,
which gives a correction of order  $(a_0/r_C)^2 \sim 10^{-6}$ times Fu's
estimate in the radiated power.
The second are retardation corrections of order $v/c$, with
$v \sim 0.3 \times  10^{-3}c$
the thermal velocity of the valence electrons, which gives a correction
of order $v^2/c^2 \sim 10^{-7}$ times Fu's estimate in the radiated power.
(There will also be a cross-term correction in the radiated power of order
$(a_0/r_C)(v/c) \sim  10^{-6}$  times Fu's value.)
Thus, taking the effect
of charge neutralization into account, Fu's calculation shows that
$\lambda$ can be at most $\sim 10^{12}$ times as large as the standard
CSL value of $\lambda$.
\bigskip
\centerline{\bf 5E.~~Heating of Protons}
\bigskip
We consider next the heating of protons by the stochastic noise, as given
by Eq.~(7).  Taking $M=m_N$, we get from Eq.~(7) and the standard
CSL parameters,
$${dE\over dt}={3\over 4}\lambda {\hbar^2\over r_C^2 m_N}
=6.8 \times 10^{-26} {\rm eV}\, {\rm s}^{-1}
=1.1 \times 10^{-37} {\rm erg}\, {\rm s}^{-1}
~~~.\eqno(12)$$
Hence over a ten billion year period ($3.15 \times 10^{17} {\rm s}$), ignoring
dissipation,
a proton would gain an energy from this effect
of $\sim 2 \times 10^{-8}$ eV,
corresponding to a temperature of $\sim 2 \times 10^{-4}$ degrees Kelvin.
If $\lambda$ were larger that the standard CSL value by a
factor of $2 \times 10^7$,
the minimum value of the range inferred from our latent image formation
analysis, the temperature increase would be $\sim 7 \times 10^3$ degrees
Kelvin over the age of the universe.

We use this number to get a number of
different bounds. The first is obtained by
supposing that if $\lambda$ is much larger than the standard value, all
of the energy gained by protons in the universe
is thermalized into low energy photons.
The total energy per unit volume released this way must not exceed a
small fraction, say 0.1, of the energy per unit volume in the 3 degree
($\sim 2.6 \times 10^{-4}$ eV)
microwave background radiation.  Taking account of the fact that there
are $\sim 10^9$ microwave background radiation photons per proton in
the universe, this
gives the bound that $\lambda$ must be less than $\sim 10^{12}$ of
the standard CSL value.

A more precise way to get a cosmological bound from proton heating is to
utilize the fact that there have been detailed experimental and theoretical
studies of low density regions of intergalactic space containing the
intergalactic medium (IGM) [22].  The IGM consists of highly ionized
hydrogen, with a typical temperature of around
$2 \times 10^4$ degrees Kelvin [22] as measured between
redshifts of $z=4$ and
$z=2$.  The IGM is heated by radiation from various astrophysical sources
(stars, supernovas, quasi-stellar objects) and is cooled by adiabatic
expansion of the universe, and by recombination cooling of the plasma.
We can get an upper bound on the proton heating rate, by ignoring possible
astrophysical heating mechanisms, and assuming that all of the heating in
fact comes from the CSL heating effect, which must be equated
to the cooling rate rate from adiabatic expansion and recombination
cooling to explain the observed temperature equilibrium.  For the highly
ionized IGM, recombination cooling is less important (by a factor of
around 6 at $z=3$) than adiabatic expansion, so to to get an estimate
we shall ignore the influence of recombination cooling, and
just consider the effect of adiabatic cooling as follows.  Under adiabatic
expansion, the temperature varies with redshift $z$ as $T \propto (1+z)^2$,
and so the thermal energy loss of a proton obeys $dE/dt=(d/dt)(3/2)kT =
\big(3kT/(1+z)\big)|dz/dt|$.  At z=3
(for Hubble constant $H_0=71$ km/s/Mpc and for
matter and dark energy fractions
$\Omega_{m}=0.26~,\Omega_{\Lambda}=0.74$) one has $|dt/dz|=0.8 \times 10^9
{\rm years}$.  Assuming a temperature $T$ at $z=3$ of
$2\times 10^4$ degrees Kelvin,
this gives an energy loss rate
of $0.5 \times 10^{-16} {\rm eV} {\rm s}^{-1}$.
Equating this to an energy
gain from proton stochastic heating, from Eq.~(12), gives a value of
$\lambda$ equal to $8 \times 10^8$ times the standard CSL value as an
upper bound.

A second IGM bound can be obtained from measurements at $z \simeq 0$, which
are less precise than those in the range $2 \leq z \leq 4$.  Ricotti,
Gnedin, and Shull [22] give a $z=0.06$ mean
temperature of $\sim 10^{3.7\pm 0.5}$
degrees Kelvin (with one sigma errors), which limits
$\lambda$ to be at most $10^{7.2\pm 0.5}$ times as large as
the standard CSL value.  Combining the $z=3$ IGM bound  obtained in the
preceding paragraph with this one, and rounding the error in the exponent
to an integer, we get an overall IGM bound on $\lambda$
of $\sim  10^{8\pm 1}$ times the standard CSL value.

Yet another cosmological bound comes from considering interstellar
dust grains.  Here we compare the rate of energy accumulation,
given by Eq.~(12), with the rate of energy radiation by the grain.
Assuming $10^{24}$ nucleons per cubic centimeter, the rate of volume
energy production for the standard CSL value of $\lambda$ is
$\sim 7 \times 10^{-2} {\rm eV} {\rm s}^{-1} {\rm cm}^{-3}$.  Radiation from
dust grains does not follow the Stefan-Boltzmann law (power $\propto
{\rm area} \times T^4$), but rather scales with the dust grain volume and
the fifth power of the temperature [23], according to the following
formula giving
$W$, the rate of energy radiated per unit volume of grain,
$$W= 32 \pi\, 24.9 {c (k T_g)^5\over (hc)^4} \kappa^{\prime}~~~.
\eqno(13)$$
Here $h$ and $c$ are the Planck constant and the velocity of light,
$T_g$ is the grain temperature, and $\kappa^{\prime}$ is the
imaginary part of
the refractive index.  Taking typical values $T_g \sim 20 $ degrees Kelvin
and $\kappa^{\prime} \sim 0.05$, this formula gives
$W\simeq 2 \times 10^{14}     {\rm eV} {\rm s}^{-1} {\rm cm}^{-3}$. Hence
the dust grain energy balance implies only a weak bound, that $\lambda$
can be at most of order $10^{15}$ times its standard CSL value.

We finally consider bounds on energy production obtained from
the energy balance associated with planetary
heat flows.  Table 6.3 of de Pater and Lissauer [24] gives the
luminosity to mass ratio $L/M$ for solar system objects.  The lowest
ratios in the table,  $<4$, $4, 6.4 $ in units of
$10^{-8} {\rm erg}\, {\rm g}^{-1} {\rm s}^{-1}$,
come respectively from Uranus, carbonaceous chondrites,
and Earth.  Comparing with the nucleon heating rate of Eq.~(12), which
is $\sim 7 \times 10^{-14} {\rm erg}\, {\rm g}^{-1} {\rm s}^{-1}$, and
ignoring
the fact that the nucleons in earth are not isolated, we would conclude
that $\lambda$ can be at most $5 \times 10^5$ times the standard CSL value.   However,
this bound is dubious, for the following reason.
In a recent paper, Bassi, Ippoliti and Vacchini [25],
following up on earlier work of Halliwell and Zoupas [26] and Gallis and
{}Fleming [27] (see also Benatti et al [28]), point out that
in stochastic models with a dissipative form,
the rate of heating of nucleons is not constant, but approaches zero after
a finite time, so that the energy gained through the heating effect
approaches a finite limit as opposed to increasing indefinitely.
Even when the fundamental stochastic reduction
process is non-dissipative, for nucleons
in the earth the reduction process is a much weaker effect (by about
28 orders of magnitude, for $\lambda$ at the IGM upper limit)
than ordinary collisional decoherence [29], which
is dissipative.  Hence any stochastic heating effect should
have equilibrated to zero very rapidly as a result of the influence of
molecular collision effects, and so Eq.~(12)
cannot be compared to the planetary heat flow to set a bound on $\lambda$.

\bigskip
\centerline{\bf 6.~~Possible modifications in our analysis}
\bigskip

We discuss in this section possible modifications in our analysis.
We consider first a modification that might eliminate the
upper bound obtained from IGM heating. The heating bound could be
eliminated if, as discussed in ref [25], the stochastic Schr\"odinger
equation itself were modified to include dissipative effects.  If the energy
gain by a proton from stochastic effects were to saturate, independent
of the value of $\lambda$, at a level significantly below $10^4$ degrees
Kelvin, then IGM heating would not lead to a bound on $\lambda$.
However, as emphasized in ref [25], for a dissipative stochastic model
to be a serious competitor
to the standard CSL model, a way has to be found to give dissipative
reduction a field theoretic formulation,
so that Fermi and Bose statistics for identical particles can be properly
taken into account, as is done in the CSL model.  This is an important
open problem needing further study, but in the absence of its solution we
continue to assume that the stochastic Schr\"odinger equation is exactly
nondissipative in form.

We consider next the effect on the heating bound of changing the
correlation function $g(x)$ and the correlation length $r_C$.  The
first comment to be made is that changing the form of $g(x)$, as long
as it remains a function only of the magnitude of $x$, has no
qualitative effect  on the analysis.
{}For example, suppose that instead of a Gaussian, one takes the simple
exponential form
$$g(x)=e^{-|x|/r_C}/(8 \pi r_C^3)~~~, \eqno(14a)$$
where the normalizing factors have been chosen to ensure that
$\int d^3x g(x)=1$.  At first sight, one might think that such a choice
would lead to the appearance of single inverse powers of $r_C$ in
various expressions, such as the small $\ell$ expansion of $\Gamma_R$
and the stochastic heating rate (see Eqs.~(9) and (7) respectively).
This, however, is not the case, because as is made clear in Eq.~(8.14)
of the review of Bassi and Ghirardi [1], the function $g(x)$ enters
the analysis only through the integral
$$G(y)=\int d^3x g(y-x) g(-x)~~~.\eqno(14b)$$
When $g(x)$ is inversion invariant \big($g(x)=g(-x)$\big),
the order $y$ term in
the Taylor expansion of $g(y-x)$ vanishes on integration over $x$, and
thus $G(y)/G(0)$ begins with an order $y^2$ term.   For example, the
choice $g(x)$ of
Eq.~(14a) leads (with $s=|y|/r_C$) to
$$\eqalign{
G(y)=& {1\over 64 \pi} (1+s+s^2/3)e^{-s}\cr
=&{1\over 64 \pi}[1-s^2/6+s^4/24-s^5/45 + {\rm O}(s^6)]~~~,\cr
}\eqno(14c)$$
in which an odd power of $|y|$  appears first at fifth order!
Hence with alternative choices of the correlation function, the qualitative
form of Eqs.~(9) and (7) remains unchanged, apart from a rescaling of
$r_C$ by a numerical constant. See Weber [1] for a further discussion of
alternative choices of the correlation function.

Continuing, then, with the Gaussian choice for $g(x)$, let us consider
the effect of changing $r_C$ from the conventional, and rather arbitrary
choice, of $r_C=10^{-5}{\rm cm}$.  Clearly, if $r_C$ were increased to
$r_C \sim  10^{-4}{\rm cm}=1 {\mu}$, the IGM heating rate given by
Eq.~(7) would be reduced
by two orders of magnitude, and the upper bound on $\lambda$  would be
raised to $10^{10\pm 1}$ times the standard CSL value.

We next analyze the effect on our latent
image bounds of increasing $r_C$.
We consider first the case of formation of a photographic latent image.
{}For a small AgBr grain of diameter $\ell= 10^{-5}{\rm cm}$, motion of the
Ag and Br ions within the grain would  be suppressed, according to
Eq.~(9), by a factor ${1\over 4} (\ell/r_C)^2 = 1/400$.  However, since
the bromine atoms that leave the grain typically move a distance of up to
a micron into the gelatine, the suppression factor for these atoms
would be only of order ${1\over 4} \int_0^1 dx x^2 =1/12$.  So with
$n=30\times 80=2400$, and as before $N=20$
and $\lambda =2.2 \times 10^{-17} {\rm s}^{-1}$,
our revised estimate from Eq.~(9) is $\Gamma_R=2.2
\times 10^{-10}{\rm s}^{-1}$,
requiring $\lambda$ to be $1.4\times 10^{11\pm 2}$ times larger than the
standard CSL value.

Turning next to latent image formation in etched track detectors,
when $r_C$ is increased by a factor of 10, the value of $n$ used in
Eq.~(9) also increases by a factor of 10, but $N$ decreases by the same
factor.  Therefore the combination $n^2 N/r_C^2$ is a factor of 10 smaller
than estimated previously, and so an increase of $\lambda$ by a factor
of $\leq 5 \times 10^{10}$ is required for the reduction rate to equal
the latent image formation rate, using attainment of chemical equilibrium
as the criterion.

To conclude this section, we note that the physiology of the eye
gives a possible argument suggesting that  $r_C$ should not be larger than
1 or 2 microns.   Consider a dense array of
photoreceptors, which one is trying to optimize {\it both} with respect
to rapidity of response, which we assume to be given by the reduction
rate $\Gamma_R$, and the ability to resolve fine detail.  If the diameter
of each photoreceptor is $\ell$, there will clearly be a tradeoff between
the two desired attributes of the array, as we vary $\ell$.  If $\ell$ is
much smaller than $r_C$, the response rate of each detector element
is suppressed by a factor of
$\ell^4$, with a factor of $\ell^2$ coming from the explicit $\ell^2$ in
Eq.~(9), and an additional factor of $\ell^2$ coming because the factor
$n$ in Eq.~(9) scales as $\ell$.  (Since we are considering here the
rate for a single photodetector element, the factor $N$ in Eq.~(9) will
be 1).  On the other hand, the spatial resolution of the array varies as
$\ell^{-2}$.  Hence one gains more in response time than one loses in
spatial resolution as $\ell$ is increased,  until $\ell$  reaches $r_C$.
Once $\ell$ is larger
than $r_C$, the increase in the response time becomes only linear in
$\ell$ rather than quartic  \big(c.f. Eqs.~(6a) and (6c), with
$n$ now a constant and $N$ increasing in proportion to $\ell$.\big) Thus
increasing $\ell$ beyond $r_C$ produces a gain in response time that
is less than the corresponding loss of spatial resolving power.  So these
considerations suggest that an optimized photoreceptor array should have
$\ell \sim r_C$.

Assuming that the evolution of the human eye has optimized
it with respect to both response time and resolving power, and using
the empirical fact [30] that the diameter of rods in the retina is $2 {\mu}$,
this reasoning suggests that the correlation length $r_C$ is unlikely to
be larger than of the order of a micron.  However, some caveats are in order.
First, since physiological
constraints prevent eucaryotic cells from being much smaller than of the
order of microns, this argument cannot be used to disfavor smaller values
of $r_C$, such as the conventionally assumed $r_C=0.1$ micron.
Also, in applying the optimized array argument to the rods of the eye,
we have implicitly assumed that with the enhanced reduction rate suggested by
latent image formation, reduction occurs directly within the individual
rods, rather than requiring molecular motions in the optic nerve as
well.  As discussed briefly in Appendix C, estimates based on the known
amplification chain in the rods support this
assumption, but are incompatible with reduction occurring in the
conformational change of an individual
rhodopsin molecule.  However, these estimates (if not overly optimistic)
suggest such a high
reduction rate in the rod amplification chain that reduction rate
is no longer a relevant factor in determining the optimized array
dimensions, which then undermines the argument just
given relating the diameter of rods to the value of $r_C$.

\bigskip
\centerline{\bf 7.~~Implications of a larger reduction rate}
\bigskip
We turn now to a discussion of experimental implications of a greatly
enhanced value of $\lambda$.  We divide our analysis into two parts,
first considering those experiments that do not rely on the secular
increase of energy associated with the CSL process, and then considering
experiments that depend on this secular energy increase (in which case
the analysis possibly would be modified in dissipative
extensions of the CSL model.)
We consider two possible choices of enhanced CSL parameters, (I) ~$r_C=
10^{-5}{\rm cm}$, $\lambda=4\times 10^{-10}{\rm s}^{-1}$, and (II)~$r_C=
10^{-4}{\rm cm}$, $\lambda=3\times 10^{-8}{\rm s}^{-1}$.
These correspond respectively to the latent image lower bounds of
$2 \times 10^7$ and $1.4 \times 10^9$ times the standard CSL value, and
are near the lower end of the respective ranges
permitted by the IGM heating upper bounds.
\bigskip
\centerline{\bf 7A.~~Experiments not based on secular energy increase}
\bigskip
\noindent  {\bf Fullerene diffraction}  For case (I),  with a grating
of $10^{-5}{\rm cm}=r_C$, setting
$\Gamma_R=\lambda n^2 =10^2{\rm s}^{-1}$
gives a value of $n\sim 5\times10^5$ at which washing
out of the diffraction pattern
would set in.  For case (II), with a grating of
$\ell=2.5\times 10^{-5}{\rm cm}
=.25 r_C$, setting $\Gamma_R=\lambda n^2 \ell^2/(4 r_C^2) =10^2
{\rm s}^{-1}$ also gives a value of $n\sim 5\times 10^5$ at which
the diffraction pattern would start to wash out.  Thus, diffraction
experiments with projectiles of molecular weight of order 500,000 (a factor
of 500 beyond what has been achieved so far) would
confront the CSL model, when the parameter values
are minimally enhanced to account for reduction in latent image formation.
\medskip
\noindent  {\bf Supercurrent decay}  For case (I), from the calculation of
refs [17] and [18] we find a supercurrent decay rate
of $ 10^{-19} {\rm s}^{-1}
\sim 1/(3 \times 10^{11}{\rm years})$,  while for case (II), we find a
supercurrent decay rate of
$8 \times 10^{-19} {\rm s}^{-1} \sim 1/(4 \times 10^{10}
{\rm years})$. A direct measurement of the supercurrent decay time constant
(as opposed to a time constant inferred from an improved resistivity
measurement based on a short measurement time) would be needed to
confront the CSL
model with enhanced parameter values.  Again, we note that these estimates
ignore recombination processes in which Cooper pairs are created.
\medskip
\noindent {\bf Mirror deflection experiment}
The papers of ref [2] give a detailed analysis of a mirror deflection
experiment proposed by Marshall et al [31], as expressed in terms
of the parameter
$\eta$ governing the small displacement CSL equation.  For the geometry
of the proposed experiment, with a cubical mirror of side $S$ and density
$D$, when $S>>r_C$ the parameter $\eta$ is given by the formula
$$\eta=8\pi r_C^2\lambda S^2 D^2~~~.\eqno(15)$$
When $S$ is of order $r_C$, this
formula has corrections of order unity, but we shall
continue to use it to give a rough order of magnitude estimate.
{}For case (I) we find that $\eta$ is a factor of $2\times 10^7$ larger than
in ref [2], giving a fringe visibility damping factor $e^{-\Lambda}$,
with $\Lambda \sim 0.04$.  The thermal decoherence background has been
estimated recently by Bern\'ad, Di\'osi and Geszti [32] as $\Lambda_T \sim
0.5$, so the case (I) CSL effect is not detectable.
{}For case (II), we find
that $\eta$ is a factor of $1.5 \times 10^{11}$ bigger than than in ref [2],
corresponding to $\Lambda \sim 3 \times 10^2$, indicating complete
suppression of the interference fringes, a large and
detectable effect.  Thus, even a version of the Marshall et al experiment
that is a factor $100$ times less sensitive than envisaged
in their proposal would be of considerable interest, provided that thermal
decoherence backgrounds can be kept small.
\bigskip
\centerline{\bf 7B.~~Experiments utilizing secular energy increase}
We discuss here experiments that utilize the fact that any non-dissipative
Brownian process, including the CSL stochastic noise, will produce a
secular increase in system energy and a corresponding increase in
the rms values of translational and rotational displacements.  Should
it be possible to formulate a satisfactory dissipative version of CSL,
the predictions for these experiments would be altered when
the observational time exceeds a characteristic time for dissipative
effects to set in (see ref [25]).   Having stated this caveat, we proceed
to give the results expected from the standard version of CSL, with
enhanced parameters as suggested by latent image formation.
\medskip
\noindent{\bf The Collett--Pearle rotational diffusion proposal}
Collett and Pearle [4] have proposed searching for the  mean
square rotational
diffusion $\Delta \theta_{\rm CSL}$ of a suspended
disk of order $r_C$ in dimensions.
We write the result of their Eq.~(C.6) in the form
$$\Delta \theta_{\rm CSL}=\left({\hbar f_{\rm ROT} I
\lambda \over 12}\right)^{1\over 2}
{t \over m_N r_C^2} \Delta \theta_{\rm SQL}~~~,\eqno(16a)$$
with $f_{\rm ROT}$ a dimensionless function of the ratios of disk dimensions
to $r_C$ given in ref [4],  with $I$ the disk moment of inertia,
and with the ``standard
quantum limit'' $\Delta \theta_{\rm SQL}$ given by
$$\Delta \theta_{\rm SQL} =\left({\hbar t\over I}\right)^{1\over 2}~~~.
\eqno(16b)$$
{}For readers not familiar with standard quantum limits, Eq.~(16b), and also
the analogous result for the standard quantum limit on measurement
of the translational coordinate of a free mass, are given in Appendix B.
\big(Note that the moment of inertia $I$ cancels
when Eq.~(16b) is substituted
into Eq.~(16a), which is why it does not appear in Eq.~(C.6) of ref [4].)
{}Following Collett and Pearle, we assume a disk radius $L=2r_C$ and a disk
thickness $b=0.5 r_C$, for which $f_{\rm ROT} \simeq 1/3$, and the
moment of inertia is given by $I=ML^2/4$, with $M$ the disk mass. Assuming
a disk density of $10^{24} {\rm cm}^{-3}$, we then
find, for case (I), $\Delta \theta_{\rm CSL}/\Delta \theta_{\rm SQL}
=6.6 \times 10^2 t$,  and for case (II)
$\Delta \theta_{\rm CSL}/\Delta \theta_{\rm SQL} =1.8 \times 10^4 t$,
with $t$ in seconds in both formulas.  For estimates of the rotational
diffusion produced by Brownian motion, which places severe pressure and
temperature constraints on the environment of the disk, see Eq.~(6.4)
of ref [4]. Since the Brownian rotational diffusion
scales as the product of the square root of the
pressure and the fourth root of the temperature, increasing the CSL effect
by a factor of order $10^3$ to $10^4$ makes the vacuum and cryogenic constraints
on the experiment much less severe.
\medskip
\noindent{\bf Nanomechanical oscillator}
We consider next the nanomechanical resonator reported by LaHaye et al
[33], and discussed from the viewpoint of CSL effects in Adler [6], to
which the reader is referred for details.
Recalculating the numbers found in ref [6] to reflect
an enhanced CSL parameter
$\lambda$ \big(and in case (II), an increased $r_C$\big), we find
that the increase in the harmonic
oscillator occupation number $N$ over a time equal to the inverse of the
noise bandwidth is of order $10^{-5}$ in case (I), and of order $10^{-1}$
in case (II).  Similarly, the root mean square
deviation of the quantum nondemolition variables $X_{1,2}$ over a time
equal to the inverse of the noise bandwidth is of order $3 \times 10^{-3}$ of the
corresponding standard quantum limit in case (I), and of order
the 0.3 times the
standard quantum limit in case (II).  Hence the enhanced CSL effect would
still be not accessible in case (I), and would come close to being
detectable in case (II).
\medskip
\noindent{\bf Gravitational wave detectors LIGO and LISA}
We consider finally gravitational wave detector experiments, also discussed
from the viewpoint of CSL effects in ref [6].  The Advanced LIGO
Interferometers [34] are expected to approach the standard quantum limit
in position measurement accuracy, while the Laser Interferometer
Space Antenna
(LISA) [35] should achieve a position accuracy of around $10^4$ times the
standard quantum limit.  Recalculating the numbers from ref [6], we find
that for LIGO the root mean square stochastic deviation in the test mass
coordinate over a time interval of 1/70 s will
be $\sim 0.02$ times the standard quantum limit in
case (I), and $\sim 1.4$ times the standard quantum limit in case (II).
{}For LISA, the corresponding figures over a time interval of $10^4$ s
are $\sim 6\times 10^4$ times the standard
quantum limit (that is $\sim 6$ times the expected position accuracy) in
case (I), and $\sim 5\times 10^6$ times the standard
quantum limit ($\sim 5\times
10^2$ times the expected position accuracy) in case (II).  Thus, the
enhanced CSL parameter values give an effect that may be barely visible
in LIGO in case (II), and should be readily
visible in LISA in both cases (I) and (II).
\medskip
\noindent{\bf The background from molecular collisions}
In designing experiments to look for CSL effects connected with the secular
increase in energy and associated coordinate deviations, one will have
to make sure that similar effects associated with molecular collisions
are smaller in magnitude.  There are two ways of estimating such effects.
The first is to use the decoherence calculation of Joos and Zeh [29]
(with later corrections [29] in the numerical coefficient),
which gives a rate parameter $\Lambda$ for decrease in the off-diagonal
density matrix element in a coordinate basis, that is analogous to the
CSL parameter $\eta/2$.  The second is to use standard classical Brownian
motion formulas, as is done in the paper of Collett and Pearle [4].
In Adler [36] we calculate and compare the mean square coordinate
fluctuation predicted by the two methods, for the case of isotropic hard
sphere scattering.  We show that the two calculations give identical results,
and so either method can be used to estimate collision backgrounds that may
mask CSL effects.

\bigskip
\centerline{\bf 8.~~Summary and discussion}
\bigskip
To summarize, we have found that if latent image formation constitutes
measurement, then the parameters of the CSL model for objective state
vector reduction must be much larger than conventionally assumed.
The enhanced parameter values needed are
compatible with empirical upper bounds, and suggest that CSL effects may
be within reach of experimental detection within the next decade or two.

{}Further experiments on latent image formation would also be of
great interest,
to firm up (or falsify) assumptions that we have made in our analysis.
It would be very useful to have more accurate information on the time
needed for a latent image to form, in both photography and etched track
detectors, since estimates of this time (or rate)
were a crucial input for our analysis.  Our central assumption, that
latent image formation, and not subsequent development or etching,
constitutes measurement, should also be subject to experimental test.
In principle, a photographic emulsion or an etched track detector could
be used as a ``which path'' detector in one arm of a quantum interferometer;
if latent image formation constitutes measurement, then the interference
fringes should be destroyed by the presence of the emulsion or detector
through formation of the latent image, which is a ``fossilized record''
of the path through the interferometer.  According to the view expressed
by Gisin and Percival [3] and adopted in this article, collapse onto a
definite outcome should occur once the latent image has formed, and should
not depend on a subsequent stage of amplification by development or
etching.
\bigskip
\centerline{\bf Acknowledgments}
I wish to thank E. Abrahams, A. Bassi, B. Barish, N. Brandt,
D. Goldhaber-Gordon, P. Goldreich,
J. Halliwell, L. Hui,
E. Ippoliti, R. Joynt, A. Levine, J. Ostriker, N. Gnedin,
L. Page, P. Pearle, G. Steigman,
M. Tinkham, H. Robins, K. Thorne, and M. Turner for
helpful conversations or email correspondence.
This work was supported in part by the Department of Energy under
Grant \#DE--FG02--90ER40542.  Part of this work was done while the
author was at the Aspen Center for Physics.
\bigskip
\centerline{\bf Appendix A: Radiation by a free charge}
\bigskip
We give here an alternative calculation of the radiation by a free charge,
obtained by using the basic dipole radiation formula
$P={1\over 6 \pi} e^2 a^2$, with $e^2/(4 \pi)=1/137.04$ and
with $a$ the acceleration of the particle of charge $e$.  For a
time average over the stochastic process, $a^2$ is to be interpreted
as $E[(\ddot x)^2]$, with $E[~]$ the stochastic expectation.  Rather than
proceeding from the CSL equations, we use a modification of the
CSL equations that has a unitary dynamics but leads to the same density
matrix evolution; this substitution is allowed because the expectation
$E[~]$ of quantum transition probabilities is expressible directly in
terms of the density matrix. (This substitution is used by Fu [21] as the
basis of
his calculation.)  The advantage of doing this is that the
modified dynamics has a Heisenberg picture formulation with a self-adjoint
Hamiltonian, which permits a direct calculation of $\ddot x$ from
the Heisenberg
equations of motion.

Specifically, corresponding to the CSL model we have the effective
coordinate representation Hamiltonian
$$H={p^2\over 2m} - {\hbar \over dt} \int d^3z (m/m_N) dB(z)
g(z-x) ~~~,\eqno(A1)$$
from which we calculate
$$\eqalign{
\dot x=&-{i\over \hbar} [x,H]={p\over m}~~~,\cr
\ddot x =&-{i\over \hbar}[p/m,H] = {\hbar\over m_N dt}  \int d^3z dB(z)
\partial_xg(z-x)~~~,\cr
}\eqno(A2)$$
with $\partial_x$ a vector gradient.
Thus for $E[(\ddot x)^2]$ we get
$$E[(\ddot  x)^2]= {\hbar^2 \gamma \over m_N^2 dt}
\int d^3z [\partial_xg(z-x)]^2
={3 \hbar^2 \lambda \over 2 m_N^2 r_c^2 dt}~~~.\eqno(A3)$$
Using $1/dt= \delta(0)=(2\pi)^{-1} \int_{-\infty}^{\infty} dk =
{\pi}^{-1}\int_0^{\infty} dk$, and substituting into the dipole radiation
formula, we get
$$ P= {e^2 \over 4 \pi^2} {\hbar^2 \lambda \over m_N^2 r_C^2}
\int_0^{\infty} dk
~~~,\eqno(A4)$$
in agreement with Eq.~(11) when specialized to  units with $\hbar=1$.
The total radiated power
is infinite, but since CSL is a nonrelativistic theory, the calculation
is reliable only for $k<<m$.  With a cutoff taken at $k=m$, the radiated
power is
smaller than the power gain through stochastic heating \big(see Eq.~(7)\big)
by a factor $e^2/(3 \pi^2)$.

We also note that exactly the same result is obtained
from the so called QMUPL model,
which corresponds to the leading small-displacement Taylor
expansion of the CSL model (and of the original Ghirardi--Rimini--Weber [37]
model as well), as
discussed and further referenced in refs [2]. In this case the effective
Hamiltonian is
$$H={p^2\over 2m} - {\hbar \over dt} (m/m_N)\surd\eta dB x ~~~,\eqno(A5)$$
with $(dB)^2=dt$ and $\eta=\lambda/(2 r_C^2)$, and one immediately finds
$$\ddot x={\hbar \over m_N} {\surd \eta \over dt} dB~~~.\eqno(A6)$$
Since the QMUPL model of Eq.~(A5) is one-dimensional, one must
multiply by a factor
of 3 in forming the expectation of the squared acceleration from Eq.~(A6),
giving the same result as obtained in Eq.~(A3) from the CSL calculation.
As one would expect, the integral over the squared gradient of $g(z-x)$
in Eq.~(A3) is just the one encountered in evaluating $\eta$ from the
second order expansion of the function $G(y)$ of Eq.~(14b), in the course
of deriving the QMUPL model from the small displacement expansion of the
CSL model.

Aside from giving a simpler derivation of Fu's result (which he obtained
by using Feynman rules to calculate the $S$ matrix for the radiation
process), the derivation given here shows that the radiation is directly
attributable to the noise-induced acceleration.  Moreover, the
radiation has a coefficient that
is independent of the particle mass $m$, because the $1/m$ in the relation
between $\dot x$ and $p$ cancels against the $m$ coming from the
noise coefficient $m/m_N$ included in the mass-proportional coupling
scheme.  This cancellation of $m$ dependence is why the radiation field from
the germanium core largely cancels that from the valence electrons.
\bigskip
\centerline{\bf Appendix B: Standard quantum limits}
\bigskip
We derive here the standard quantum limit for measurement of the angle
$\theta$ through which a suspended disk rotates about the axis of
suspension, taken here as the $z$ axis. We then state by analogy
the corresponding standard quantum limit for the measurement
of a translational coordinate.

Let $L_z=-i\hbar \partial/\partial \theta$
be the component of angular momentum around the $z$ axis, which obeys
the commutation relation
$$[\theta,L_z]=i\hbar~~~.\eqno(B1)$$
We shall only consider very small angular displacements, and so we neglect
complications associated with the $2\pi$ periodicity of $\theta$ and the
corresponding integer quantization of $L_z$.  From the commutation relation
of Eq.~(B1), one gets an uncertainty relation of the usual form,
$$\Delta \theta \Delta L_z \geq {\hbar \over 2}~~~.\eqno(B2)$$
Suppose at $t=0$ a measurement of $\theta$ is made to accuracy $\Delta
\theta$.  This induces an uncertainty
$\Delta L_z \geq \hbar/(2 \Delta \theta)$,
and since $\Delta L_z =I \Delta \omega_z$, with $I$ the moment of inertia
of the disk around the $z$ axis and  $\omega_z$ the corresponding angular
velocity, the squared uncertainty in the angular coordinate of the disk
at time t is
$$\big(\Delta \theta(t)\big)^2=(\Delta \theta)^2 + (t \Delta \omega_z)^2
=(\Delta \theta)^2 + (t \hbar)^2 /(2 I \Delta \theta)^2~~~.\eqno(B3)$$
Minimizing this expression with respect to $\Delta \theta$, we get the
standard quantum limit on the measurement accuracy of $\theta$ at time $t$,
$$\big(\Delta \theta(t)\big)^2 \geq (\Delta \theta_{\rm SQL})^2~,~~
\Delta \theta_{\rm SQL}=(\hbar t/I)^{1/2}~~~.  \eqno(B4)$$

In an entirely analogous fashion, from the commutation relation
$[x,p]=i\hbar$, together with $p=Mv$ with $M$ the mass and $v$ the velocity,
one derives the standard quantum limit for measurement of a translational
coordinate, $\Delta x_{\rm SQL}= (\hbar t/M)^{1/2}$.  For further details of
standard quantum limits, see ref [38].
\bigskip
\bigskip
\centerline{\bf Appendix C: Where does reduction occur in the visual
system?}
\bigskip
We give here some estimates for the implications of an enhanced $\lambda$
parameter for the human visual system. Attention was first drawn to
the issue of possible state vector reduction in the visual system
by Albert and Vaidman [39], with a response in Aicardi et al [39], as
reviewed in Bassi and Ghirardi [1]. An issue raised in this early work
is: where does one expect reduction  to occur in the visual system  --
at the retina, or higher up in the nervous system or brain?

To address this question, we shall assume case (I) discussed
in Sec. 7, that is $r_C=10^{-5} {\rm cm}$ and $\lambda=4 \times 10^{-10}
{\rm s}^{-1}$.  We first ask whether the cis-trans conformational change
of an individual rhodopsin molecule on absorption of a photon can give
reduction with these parameter values. A rhodopsin molecule has a molecular
weight of about $4 \times 10^4$ nucleons, and a diameter of about $4 \times
10^{-7} {\rm cm}$.  Assuming that in the conformational change all of the
nucleons in the molecule move by a molecular diameter (likely a considerable
overestimate) we find from Eq.~(9), with $n=4 \times 10^4$,
$N=1$, and $\ell/(2 r_C)=2 \times 10^{-2}$, a reduction rate of $\sim
3 \times 10^{-4} {\rm s}^{-1}$.  In other words, reduction with these
parameters requires of order $3\times 10^3{\rm s}$, while the
conformational change occurs
in $200{\rm fs}= 2\times 10^{-13}{\rm s}$, so even with the enhanced
$\lambda$ parameter, reduction cannot occur during the conformational change.
{}For a detailed discussion
of the possibility of reduction during conformational change, and many
useful references, see Thaheld [40].

On the other hand, with the enhanced $\lambda$ parameter, there is no
apparent difficulty in reduction occurring during the amplification chain
in a rod cell, without invoking signal transport in the nervous system to
which the rod is linked.  According to the review of Rieke and Baylor
[41], a single catalytically active rhodopsin leads to the closure of
several hundred ion channels, which over the response time of $300 {\rm ms}$
blocks the entrance of about 3000 cations (sodium or potassium) to the
outer segment of the rhodopsin molecule.  Assuming that
the relevant $\ell$ here
is of the order of the rod cell diameter of $2~ {\rm microns}
>r_C$, we can use the estimate of Eq.~(8), with $N=1$ and $n=300 \times 3000
\times 23 \sim 2\times 10^7$, giving a reduction rate of $2\times 10^5
{\rm s}^{-1}$. Even if our assumptions here
about $n$ and $\ell$ are optimistic,
this estimate indicates that with the enhanced $\lambda$ parameter, there
is plenty of latitude for reduction to be
complete by the end of the amplification chain in the individual rod cell.
Note that with the original CSL $\lambda$
parameter value, the reduction rate would be too small by about three orders
of magnitude for reduction to occur in a rod cell, requiring the
invocation of signal transport in the nervous
system, as discussed in ref [1].
\bigskip
\centerline{\bf Appendix D:  Reduction arising from phonon emission
and thermal fluctuations}
\bigskip
In this appendix we estimate the reduction rate associated with phonon
emission by electrons and ions, and also the reduction rate associated
with thermal lattice fluctuations.

To start, we note that phonons of wavelength shorter than $r_C$
lead to no net center of mass displacement of a block of emulsion of
linear dimensions $r_C$, and so are not the main phonon contributors to
reduction.  On the other hand, long wavelength phonons can lead to
a displacement of the center of mass of such a block, and so can potentially
have a significant effect.  Thus, we are interested in making an estimate
of the reduction effect of the emission of long wavelength acoustic phonons.
The effect of short wavelength optical phonons will be taken into account
afterwards when we estimate the reduction rate associated with thermal
lattice fluctuations.

Consider, then, a lattice block of total mass $M$ with atom
site  coordinates $u_i$.  For a phonon of
angular frequency $\omega$, each site
will oscillate as $u_i=a\cos (\omega t +\delta_i)$, with $\delta_i$ a
site-dependent relative phase, and so the time-averaged velocity is
$\langle u_i^2 \rangle= a^2 \omega^2/2 \equiv \omega^2 \ell^2$,
independent of $i$.
Since the phonon energy
is half kinetic and half potential, we thus have
$${1\over 2} M \omega^2
\ell^2 = {1\over 2} \hbar \omega~~~,\eqno(D1a)$$
giving for the mean square atomic displacement $\ell^2$,
$$\ell^2 = {\hbar \over M \omega}~~~.\eqno(D1b)$$
For a block composed of $N$ sub-blocks of dimension of the correlation
length $r_C$, each containing $n$ nucleons, we have $M=Nnm_N$,
while the reduction rate is given by Eq.~(9) in terms of $N,n,\ell$.
Substituting Eq.~(D1b) we get
$\Gamma_R=\lambda f(\omega)$, with
$$f(\omega)={\hbar n \over 4 r_C^2 m_N \omega}~~~.\eqno(D2a)$$
Note that the factor $N$ has dropped out, because when the lattice
contains many
coherent length sized blocks, the $N$ in Eq.~(9) cancels against the
$1/N$ implicit in Eq.~(D1b), which reflects the fact that the energy of
the single phonon is dispersed over over the whole lattice, and so the
mean squared amplitude of oscillation of each atomic site scales as  $1/N$.
Note also that Eq.~(D2a) still assumes that the phonon
wavelength is longer than $r_C$, so that all nucleons within a block of
size $r_C$ move together, and
the $n^2$ factor in Eq.~(9) is applicable.  To account for the loss of
coherence when the phonon wavelength is shorter than $r_C$, a factor
$$G(\omega) = {\rm min}[1, (\lambda/r_C)^3] =
 {\rm min}[1, \big(2 \pi c_s /(r_C\omega)\big)^3] ~~~\eqno(D2b)$$
must be included in Eq.~(D2a), with $c_s$ the acoustic phonon velocity (the
velocity of sound, typically $c_s \sim 3 \times 10^5 {\rm cm}\, {\rm s}^{-1}$).
This factor takes the value of unity for $\lambda \geq r_C$, and approaches
$1/n$ as $\lambda$ approaches the lattice spacing $\sim 10^{-8} {\rm cm}$.
We will proceed by first doing our estimates without including $G$,
and then noting the correction factor, denoted by $\langle G \rangle$,
when $G$ of Eq.~(D2b) is included in the integrands.

Consider now an electron with initial wave number $k_0$ that
slows down through multiple emissions of acoustic phonons, as described in
Kittel [42].  We wish to integrate Eq.~(D2a) over the spectrum of
emitted phonons, to get the total reduction rate $\lambda f_{ \rm tot}$
attributed to the emitted phonons. Letting $R\big(k(t),q\big)$ be the
rate of production of phonons of wave number $q$ when the electron wave
number is $k(t)$, we have
$$f_{\rm tot}=\int_0^T dt \int_0^{q_{\rm max}} dq R\big(k(t),q\big)
f\big(\omega(q)\big) ~~~.\eqno(D3a)$$
The upper limit $T$ of the time integration is determined by
$k(T)=k_{\rm th}$, where $k_{th}=(3 m^* k_B T)^{1\over 2}/\hbar$
is mean thermal electron wave number
(with $m^*$ the electron effective mass),
since once the electron has been thermalized by phonon emission, it
is in an equilibrium where phonon absorption by the electron is as important
as phonon emission.  We shall assume that both $k_0$ and $k_{\rm th}$ are
much larger than the threshold momentum for emitting acoustic phonons, given
in terms of $m^*$ and $c_s$ by $k_{\rm min}= m^*c_s/\hbar$.
With this assumption (which we shall verify), the upper limit $q_{\rm max}$
giving the maximum phonon wave number that can be emitted by an electron of
wave number $k$ is given by $q_{\rm max}=2k$.   Also with this
assumption, the time evolution of the
electron wave number $k$ is given by $dk/dt=-\sigma k^3$, with $\sigma$
a constant that will end up dropping out of the calculation. This allows
us to change variables from $t$ to $k$, and so Eq.~(D3a) becomes
$$f_{\rm tot}=\int_{k_{\rm th}}^{k_0} dk \sigma^{-1} k^{-3}  \int_0^{2k} dq
R(k,q) f(c_s q) ~~~,\eqno(D3b)$$
where in the argument of $f$ we have substituted $\omega=c_s q$.

The analysis of Kittel [42] also gives an expression for $R(k,q)$,
$$R(k,q)={5 \over 16} \sigma {q^2 \over k}~~~,\eqno(D4)$$
with $\sigma$ the same constant that appears in the time evolution of
the electron wave number. So substituting Eq.~(D4) into Eq.~(D3b), and
doing the integrals, we get
$$f_{\rm tot}={5 \over 32} {\hbar n \over r_C^2 m_N c_s}
\left({1\over k_{\rm th}} - {1 \over k_0} \right)
\simeq {5 \over 32} {\hbar n \over r_C^2 m_N c_s k_{\rm th} }
~~~;\eqno(D5a)$$
note that this result is essentially independent of the initial electron
wave number $k_0$, as long as this is significantly greater than the
thermal wave number $k_{\rm th}$.
Taking the electron effective mass $m^*$ equal to the electron mass,
we get (at room temperature) $k_{\rm th}=  10^7 {\rm cm}^{-1}$,
while for the minimum wave number for phonon emission we get
$k_{\rm min}= 0.3 \times 10^6 {\rm cm}^{-1}$, which is much smaller than
$k_{\rm th}$, as assumed.  Evaluating Eq.~(D5a) with this value of
$k_{\rm th}$, with $r_C=10^{-5} {\rm cm}$ and $n= 10^9$, we
get $f_{\rm tot}\simeq 0.3 \times 10^3$, giving $\Gamma_R=\lambda f_{\rm tot}
\sim  10^{-14} {\rm s}^{-1}$, which is a factor of $10^6$ smaller
than the latent image formation estimate of Eq.~(8).  This is still an
overestimate, since we have not included the factor $G$ of Eq.~(D2b);
when the integrals are redone with the factor of $G$ included, we find
an additional factor in Eq.~(D5a) of
$$\langle G \rangle=  \left( { \pi \over r_C k_{\rm th} } \right)^2
\simeq  10^{-3}~~~.\eqno(D5b)$$ Thus the reduction rate induced by
phonon emission from an electron emitted from a silver halide
molecule is more than a factor of $10^9$ smaller than Eq.~(8).

The silver and bromine ions that diffuse to the surface of the grain
will already be
in thermal equilibrium.  We assume that they obey a dispersion relation of
the form $E=p^2/(2m^*)$, with $m^*$ an effective mass similar to the ionic
mass,
and so their mean wave numbers will be  given in terms of this effective mass
by
$k_{\rm th}=(3 m^* k_B T)^{1\over 2}/\hbar$.   For a particle
of effective mass $m^*$
the ratio $k_{\rm min}/k_{\rm th}$ scales as $(m^*)^{1\over 2}$, and so
for an
ion of atomic weight $\sim 100$ we get by comparison with the electron case
calculated above,
$$k_{\rm min}/k_{\rm th}
\sim (2 \times 10^5)^{1\over 2} \times 3 \times 10^{-2} \sim 10~~~.
\eqno(D6)$$
Hence only the extreme tail of the thermal velocity distribution for these
ions is above threshold for phonon production, leading to a suppression of
the associated reduction rate by a factor
$\exp(-1.5 k_{\rm min}^2/k_{\rm th}^2)\sim 10^{-65}$.

When both electrons and interstitial
ions are in thermal equilibrium, they continue to emit and
absorb phonons. The associated reduction rate will be driven by the thermal
average of
$\sum_i (u_i^1-u_i^2)^2$, with $u_i^{1,2}$ the atomic coordinates in
superimposed states in the wave function labeled by the respective indices
1,2.  An upper bound on this sum should be given by ignoring expected
cancellations in the individual terms (that is, for most atoms we expect
$u_i^1 \simeq u_i^2$), and replacing the sum by twice the thermal average of
$\sum_i u_i^2$.  The associated reduction rate is given in turn by the twice
the integral of
Eq.~(D2a), including the factor $G$ of Eq.~(D2b), over the Debye phonon
spectrum.  When $G=1$, the result is
$\Gamma_R=2 \lambda f_{\rm tot}$, where
$$f_{\rm tot}= {9\over 4} {k_B T n^2 N \over r_C^2 M_{\rm atom} \omega_D^2}
~~~,\eqno(D7a)$$
with $\omega_D$ the Debye frequency. [A similar result, with $\omega_D^2$
replaced by $3 \omega_{\rm opt}^2$, with $\omega_{\rm opt}$ the lattice
vibration frequency, is readily obtained from Eq.~(9), when $\ell^2$ is
taken as the thermal average of the atomic position, given by
$\langle u^2 \rangle = 3 k_B T/(M_{\rm atom} \omega_{\rm opt}^2)$.]
Including the coherence factor $G$ in the integral over the Debye spectrum
gives an extra factor
$$\langle G \rangle = 3 \pi c_s / (r_c \omega_D)~~~.\eqno(D7b)$$
Evaluating Eqs.~(D7a) and (D7b) at room temperature, with $r_C=10^{-5}
{\rm cm}$, $n=10^9$, $M_{\rm atom}=100 m_N$, and $\omega_D=3 \times
10^{13} {\rm s}^{-1}$, we get
$$\Gamma_R \sim 2 \times 10^{-9} N {\rm s}^{-1}~~~, \eqno(D8a)$$
or with $N=20$,
$$\Gamma_R \sim 4 \times 10^{-8} {\rm s}^{-1}~~~,\eqno(D8b)$$
a factor of three bigger than the latent image formation estimate
of Eq.~(8).  As noted above, Eq.~(D8b) is expected to give  a substantial
overestimate
of the contribution of thermal fluctuations to the reduction rate, since
for most atoms one expects $(u_i^1-u_i^2)^2 << (u_i^1)^2+(u_i^2)^2$.

\vfill\eject
\centerline{\bf References}
\bigskip
\noindent
[1] For reviews and references, see Bassi A and Ghirardi G C 2003
Dynamical reduction models {\it Phys. Rep.} {\bf 379} 257;
Pearle P 1999 Collapse models {\it Open Systems and Measurements in
Relativistic Quantum Field Theory (Lecture Notes in Physics} vol 526)
ed H-P Breuer and F Petruccione (Berlin: Springer).  See also Ref [6] of
Bassi A, Ippoliti I and Adler S L 2005 {\it Phys. Rev. Lett.} {\bf 94}
030401, and Adler S L 2004 {\it Quantum Theory as an Emergent Phenomenon}
(Cambridge: Cambridge University Press) ch 6. Alternative choices of
the correlation function are discussed in Weber T (1990)
{\it Nuovo Cimento B} {\bf 106} 1111. \hfill \break
\bigskip
\noindent
[2] Bassi A, Ippoliti I and Adler S L 2005 {\it Phys. Rev. Lett.} {\bf 94}
030401;  Adler S L, Bassi A and Ippoliti I  2005 {\it J. Phys. A: Math-Gen}
{\bf 38} 2715; Adler S L 2005 {\it J. Phys. A: Math-Gen} {\bf 38} 2729.
{}For a review of tests of quantum mechanics from a viewpoint not specifically
focused on dynamical reduction models, see Leggett A J 2002
{\it J. Phys.: Condens Matter} {\bf 14} R415.
\hfill\break
\bigskip
\noindent
[3] Gisin N and Percival I C 1993 {\it J. Phys. A: Math-Gen} {\bf 26} 2245.
\hfill\break
\bigskip
\noindent
[4] Collett B and Pearle P 2003 {\it Found. Phys.} {\bf 33} 1495.
See also Collett B, Pearle P, Avignone F and Nussinov S 1995
{\it Found. Phys.} {\bf 25} 1399; Pearle P, Ring J, Collar J and
Avignone F 1999 {\it Found. Phys.} {\bf 29} 465; Pearle P 2005
{\it Phys. Rev. A} {\bf 71} 032101.
\hfill\break
\bigskip
\noindent
[5] Pearle P and Squires E 1994 {\it Phys. Rev. Lett.} {\bf 73} 1.
The parameter $1/T$ of Pearle and Squires, when their $m_0$ is taken as
the nucleon mass, corresponds to the parameter $\lambda$ used here.
With this identification, their Eq. (10) gives Eq.~(7) above.
\hfill\break
\bigskip
\noindent
[6] Adler S L 2005 {\it J. Phys. A: Math. Gen.} {\bf 38} 2729. This paper
gives the energy loss rate in terms of a parameter $\eta$, that for a
nucleon is given by $\eta=\gamma/(16 \pi^{3/2}r_C^{5})= \lambda/(2r_C^2)$.
Eq.~(12) of Adler [6], and the corresponding Eq.~(6.58) of Bassi and Ghirardi [1],
both give the energy loss rate in one dimension, and so must
be multiplied by 3 to give Eq.~(7) above.
\hfill\break
\bigskip
\noindent
[7] Mott N F and  Gurney R W 1940 {\it Electronic Processes in Ionic
Crystals} (Oxford: Oxford University Press) ch VII.  See pp 232-233
for estimates of the accretion rate of silver atoms, and Table 28 giving
examples.  See also p 248, which cites
an experimental result of Berg suggesting a step time of
$1/25,000=4\times 10^{-5}$ s for the ionic stage of
latent image formation.\hfill\break
\bigskip
\noindent
[8] Avan L, Avan M, Blanc D and Teyssier J-L 1973 {\it Ionographie:
\'Emulsions -- D\'etecteurs Solides de Traces} (Paris: Doin Editeurs).
{}Fig. II.1 on p 75 gives a descriptive rendition of the Gurney-Mott process,
and grain spacing estimates of 1.6 to 5 $\mu$ are given on p 102.
Table VIII.1 on p 268 gives estimates of etchable track diameters, and
Table VIII.3 on p 278 gives times $t$ for various stages of the track
formation process.
\hfill\break
\bigskip
\noindent
[9] Berg W F 1946-7 {\it Reports on Progress in Physics} {\bf XI} 248. Size
of grains is given on p 249, along with a spacing estimate of $1 {\mu}$;
an estimate of 30 silver atoms to produce a
latent image speck is given on p 272; an ionic  step time of
$4\times 10^{-5}$ s is given on p 290.  \hfill\break
\bigskip
\noindent
[10] Hamilton J F and  Urbach F 1966 The Mechanism of the Formation of the
Latent Image  {\it The Theory of the Photographic Process, 3rd Edition}
ed C E K Mees and T H James
(New York: Macmillan and London: Collier-Macmillan). An estimate of 3 to
6 silver atoms for 50\% development probability is given on p 102;  Fig. 5.4,
showing the extent of halogen diffusion into the gelatine, is on p 94.
\hfill\break
\bigskip
\noindent
[11] Gurney R W and Mott N F 1938 {\it Proc. Roy. Soc.} {\bf 164} 151.
\hfill\break
\bigskip
\noindent
[12] Durrani S A and  Bull R K 1987 {\it Solid State Nuclear Track
Detection} (Oxford and New York: Pergamon).  See pp 34, 37, and 38 for a
discussion of latent track geometry, and p 44 for the minimum time
for a track to form.\hfill\break
\bigskip
\noindent
[13] Katz R and Kobetich E J (1968) {\it Phys. Rev.} {\bf 170} 401.
Information on the $\delta$-ray energies is given on p 403.\hfill\break
\hfill\break
\bigskip
\noindent
[14] Arndt M, Nairz O, Vos-Andreae J, Keller C, van der Zouw G and
Zeilinger A 1999  {\it Nature} {\bf 401} 680.\hfill\break
\bigskip
\noindent
[15] Nairz O, Arndt M and Zeilinger A 2000 {\it J. Mod. Optics}
{\bf 47} 2811. \hfill\break
\bigskip
\noindent
[16] Nairz O, Brezger B , Arndt  M and Zeilinger A 2001 {\it Phys. Rev.
Lett.} {\bf 87} 160401. \hfill\break
\bigskip
\noindent
[17] Buffa M, Nicrosini O and  Rimini A 1995 {\it Found. Phys. Lett.}
{\bf 8} 105.\hfill\break
\bigskip
\noindent
[18] Rimini A 1995 Spontaneous Localization and Superconductivity
{\it Advances in Quantum Phenomena} ed  E Beltrametti and J-M L\'evy-Leblond
(New York and London: Plenum Press). \hfill\break
\bigskip
\noindent
[19] Rae A I M 1990 {\it J. Phys. A: Math Gen.} {\bf 23} L57. \hfill\break
\bigskip
\noindent
[20] Tinkham M 1996 {\it Introduction to Superconductivity} 2nd ed
(New York: McGraw Hill) p 2.  For original papers, see:   Quinn  D J and
Ittner W B  1962 {\it J. Appl. Phys.} {\bf 33} 748;  Broom R F 1961
{\it Nature} {\bf 190} 992;  Collins S C, as quoted in Crowe J W 1957
{\it IBM J. Research Develop.} {\bf 1} 294.  The best direct measurement
of the time constant for supercurrent decay gives $\sim 250$ years; the
estimate $10^5$ years is obtained by scaling this result up using the
results of resistivity measurements on thin film superconductors,
taken over periods of 3 to 7 hours.  \hfill\break
\bigskip
\noindent
[21] Fu Q 1997 {\it Phys. Rev. A} {\bf 56} 1806.\hfill\break
\bigskip
\noindent
[22]  Samui S, Subramanian K and Srianand R 2005 arXiv:astro-ph/0505590,
opening paragraph, compilation of data,
and references cited; Hui L and Haiman Z 2003
{\it Ap. J.} {\bf 596} 9;
Theuns T et al 2002 {\it Mon. Not. Roy. Astron. Soc.} {\bf 332} 367;
Ricotti M, Gnedin N and Shull J 2000 {\it Ap. J.} {\bf 534} 41, table 7;
Hui L and Gnedin N 1997 {\it Mon. Not.
Roy. Astron. Soc.} {\bf 292} 27;  Spitzer L 1978 {\it Physical Processes
in the Interstellar Medium} (New York, John Wiley), ch 6, p 143;
Dalgarno A and
Mc Cray R (1972) {\it Ann. Rev. Astron. Astroph.} {\bf 10}, 375, p 383.
Note that the recombination cooling rate curves graphed in the last two
references apply only when the degree of ionization is low; when the IGM
is highly ionized, recombination cooling is several orders of magnitude
smaller than suggested by these curves, and is in fact small relative to
adiabatic cooling.  I wish to thank Nick Gnedin for a conversation which
clarified this point.
\hfill\break
\bigskip
\noindent
[23]  Kaplan S A and Pikelner S B  1970 {\it The Interstellar Medium}
(Cambridge, MA: Harvard University Press) pp 211-212.\hfill\break
\bigskip
\noindent
[24]  de Pater I and Lissauer J J 2001 {\it Planetary Sciences}
(Cambridge: Cambridge University Press) pp 224-225. \hfill\break
\bigskip
\noindent
[25] Bassi A, Ippoliti E and Vacchini B 2005 {\it J. Phys. A: Math-Gen}
{\bf 38} 8017.\hfill\break
\bigskip
\noindent
[26] Halliwell J and Zoupas A 1995 {\it Phys. Rev. D} {\bf 52} 7294.
\hfill\break
\bigskip
\noindent
[27]  Gallis M R  and Fleming G N {\it Phys. Rev. A} {\bf 43} 5778.
\hfill\break
\bigskip
\noindent
[28] Benatti F, Ghirardi G C, Rimini A and Weber T 1988 {\it Nuovo Cimento B}
{\bf 101} 333.\hfill\break
\bigskip
\noindent
[29] Joos E and Zeh H D 1985 {\it Z. Phys. B} {\bf 59} 223.
See Eqs.~(3.58), (3.59), and Table 2. Corrections to the analysis of
Joos and Zeh are given in Gallis M R and Fleming G N (1990)
{\it Phys. Rev. A} {\bf 42} 38; Dodd P J  and Halliwell J  J (2003)
{\it Phys. Rev. D}
{\bf 67} 105018; Hornberger K and Sipe J E (2003) {\it Phys. Rev. A} {\bf 68}
012105. \hfill\break
\bigskip
\noindent
[30] Davson H 1980 {\it Physiology of the Eye} 4th ed (New York \&
San Francisco:
Academic Press) p 167. \hfill\break
\bigskip
\noindent
[31] Marshall W, Simon C, Penrose R and Bouwmeester D 2003
{\it Phys. Rev. Lett.} {\bf 91} 130401.\hfill\break
\bigskip
\noindent
[32] Bern\'ad J Z, Di\'osi L and Geszti T
arXiv: quant-ph/0604157.\hfill\break
\bigskip
\noindent
[33] LaHaye M D, Buu O, Camarota B and Schwab K C 2004 {\it Science}
{\bf 304} 74.\hfill\break
\bigskip
\noindent
[34] Abramovici A et al 1992 {\it Science} {\bf 256} 325;
Barish B C and Weiss R 1999 {\it Phys. Today} {\bf 52} 44;
Shawhan P S {\it Am Sci} {\bf 92} 350.\hfill\break
\bigskip
\noindent
[35] Alberto J 2004 LISA {\it Preprint} gr-qc/0404079;
Irion R 2002 {\it Science} {\bf 297} 1113.\hfill\break
\bigskip
\noindent
[36] Adler S L (2006) arXiv: quant-ph/0607109.\hfill\break
\bigskip
\noindent
[37]  Ghirardi G C, Rimini A and Weber T 1986 {\it Phys. Rev. D} {\bf 34}
470.\hfill\break
\bigskip
\noindent
[38] Braginsky V B and Khalili F Ya 1992 {\it Quantum Measurements}
(Cambridge: Cambridge University Press); Caves C M, Thorne K S, Drever R W,
Sandberg V D and Zimmerman M 1980 {\it Rev. Mod. Phys.} {\bf 52}, 341.
I have used the definition of standard quantum limit given in Eq.~(3.2) of
the Caves et al article; the definition of Braginsky and Khalili is
a factor of $\surd 2$ smaller.
\hfill\break
\bigskip
\noindent
[39] Albert D Z and Vaidman L (1989) {\it Phys. Lett. A} {\bf 139} 1;
Aicardi F, Borsellino A, Ghirardi G C and Grassi R (1991)
{\it Found. Phys. Lett.} {\bf 4} 109. \hfill\break
\bigskip
\noindent
[40] Thaheld F (2005) arXiv: quant-ph/0509042 and
quant-ph/0604181.\hfill\break
\bigskip
\noindent
[41] Rieke F and Baylor D (1998) {\it Rev. Mod. Phys.} {\bf 70} 1027.
\hfill\break
\bigskip
\noindent
[42] Kittel C (1996) {\it Introduction to Solid State Physics} 7th ed
(New York: John Wiley \& Sons) Appendix J pp 662-665.\hfill\break
\vfill
\eject
\bye